\documentclass{aa}  
\usepackage{graphicx}
\usepackage{hyperref}
\usepackage{txfonts}
\usepackage{natbib}
\usepackage[bottom]{footmisc}
\bibpunct{(}{)}{;}{a}{}{,}

\usepackage{amstext}
\usepackage{color}

\begin{document} 

\bibliographystyle{aa} 

   \title{Distinguishing the albedo of exoplanets from stellar activity}
   \titlerunning{Distinguishing the albedo of exoplanets from stellar activity}
   \subtitle{}

   \author{L. M. Serrano\inst{1,2},
          S. C. C. Barros\inst{1},
          M. Oshagh\inst{1,3},
          N. C. Santos\inst{1,2},
          J. P. Faria\inst{1,2}, \\
                  O. Demangeon\inst{1},
          S. G. Sousa\inst{1}
          \and
                  M. Lendl\inst{4}
          }
   \authorrunning{L. M. Serrano et al.}
   \institute{Instituto de Astrof\'isica e Ci\^encias do Espa\c{c}o, Universidade do Porto, CAUP, Rua das Estrelas, PT4150-762 Porto, Portugal\\
              \email{luisa.serrano@astro.up.pt}
         \and
              Departamento\,de\,Física\,e\,Astronomia,\,Faculdade\,de\,Ciências,\,Universidade\,do\,Porto,\,Rua\,Campo\,Alegre,\,4169-007\,Porto,\,Portugal
         \and              
              Institut für Astrophysik, Georg-August-Universität Göttingen, Friedrich-Hund-Platz 1, 37077 Göttingen, Germany
         \and
              Space Research Institute, Austrian Academy of Sciences, Schmiedlstr 6, 8042 Graz, Austria
             }

   \date{Received ...; accepted ...}

  \abstract
   {Light curves show the flux variation from the target star and its orbiting planets as a function of time. In addition to the transit features created by the planets, the flux also includes the reflected light component of each planet, which depends on the planetary albedo. This signal is typically referred to as phase curve and could be easily identified if there were no additional noise. As well as instrumental noise, stellar activity, such as spots, can create a modulation in the data, which may be very difficult to distinguish from the planetary signal.}
   {We analyze the limitations imposed by the stellar activity on the detection of the planetary albedo, considering the limitations imposed by the predicted level of instrumental noise and the short duration of the obervations planned in the context of the CHEOPS mission.}
   {As initial condition, we have assumed that each star is characterized by just one orbiting planet. We built mock light curves that included a realistic stellar activity pattern, the reflected light component of the planet and an instrumental noise level, which we have chosen to be at the same level as predicted for CHEOPS. We then fit these light curves to try to recover the reflected light component, assuming the activity patterns can be modeled with a Gaussian process.}
   {We estimate that at least one full stellar rotation is necessary to obtain a reliable detection of the planetary albedo. This result is independent of the level of noise, but it depends on the limitation of the Gaussian process to describe the stellar activity when the light curve time-span is shorter than the stellar rotation. As an additional result, we found that with a $6.5$ magnitude star and the noise level of CHEOPS, it is possible to detect the planetary albedo up to a lower limit of $R_p = 0.03 \; R_*$. Finally, in presence of typical CHEOPS gaps in the simulations, we confirm that it is still possible to obtain a reliable albedo.}
        {}

   \maketitle
   
%

\section{Introduction}

   The search for extrasolar planets is a very active subject of astrophysics today. The number of confirmed exoplanets increases every day (see the updated list at http://exoplanet.eu/catalog/), and their characterization is improving through more precise data analysis techniques and the higher precision of the instruments. Some of the current aims of this field are now centered around different objectives. One of them is the identification of an Earth sibling, a planet with a similar mass and size that is situated in the habitable zone of its star \citep[e.g.,][]{Vladilo1, Gillon, Ribas, Turbet, Omalley}. Another objective is the detection of the atmospheric signature of exoplanets.
   
   From the planetary primary transit, it is possible to derive the transmission spectrum of a planet: different transit depths in different wavelengths imply that the planetary radius varies
with the wavelength \citep{Burrows, Kirk, Sedaghati}, and they
imply different interactions of the atmosphere with the stellar flux. Significant peaks at certain wavelengths are justified with stronger absorptions by certain molecules \citep[e.g.,][]{Howe, Kreidberg, Deming, Charbonneau2}. Furthermore, the emission spectrum can be obtained by analyzing the decrement of the flux during the secondary transit as a function of the wavelength \citep[e.g.,][]{Line, Haynes, Knutson2}. Another technique for probing the emission and reflected spectrum requires the identification of Doppler variations induced on spectral lines by a transiting \citep{Birkby} or even a non-transiting planetary companion \citep[e.g.,][]{Brogi, Martins}.
   
   Alternatively, it is possible to glean information about the atmosphere of a planet by analyzing the photometric phase curve, which contains the variation in planetary flux as a function of the orbital phase. For a homogeneously emitting star, the phase curve would show the primary transit (when the planet passes in front of the stellar disk) and the secondary transit (when the planet is hidden by the star), as well as three additional modulations. The first modulation is the Doppler boosting effect; it consists of a modification of the stellar brightness proportional to the radial velocity variation induced by the planet \citep{Barclay}. The second effect is the ellipsoidal modulation: by the gravitational attraction of the planet on the stellar surface, the star is deformed, with a surface tide that moves following the planet \citep{Esteves}. The last component is the planetary flux, which mainly contains atmospheric emission if the observations are performed in the infrared, and atmospheric reflection at optical wavelengths. The reflected light depends on the planetary albedo. For the hottest planets, the thermal emission might dominate the light coming from the planet, especially in the case in which the photometric observations cover a wavelength range that reaches far into the red (e.g., Kepler planets). These cases are not taken into account for the present paper since we focus on the planetary flux that is dominated by the atmospheric reflection.
   
   Unfortunately, instrumental noise and stellar activity can prevent us from identifying these different effects and therefore prevent a possible characterization of the planetary atmosphere \citep{Oshagh4}. While the instrumental noise can be significantly reduced with the adoption of high-precision instruments, stellar activity cannot be avoided in general. For instance, \citet{Gilliland} analyzed the relative contribution of instrumental noise and stellar intrinsic noise to the overall noise of Kepler stars and showed that the largest contribution comes from stellar-induced noise (see, however, \citet{Basri}). 

   Previous works have studied real stellar light curves, modeling both the primary and the secondary transit and also the beaming effect, ellipsoidal modulation, and the reflected light component. \citet{Angerhausen} have performed a study on a large sample of Kepler stars known to host hot Jupiters, with the aim to identify secondary transits and planetary albedo values. The selected stars were generally photometrically quiet. Unfortunately, not all stars are as quiet as those of this sample. For instance, \citet{Basri} have performed a comparison of the activity level of Kepler stars with that of the active Sun, and they have shown that $30 \%$ of the stars in the sample are more active than the Sun. 
   
   The present paper aims at determining the limits imposed by the instrumental noise and the stellar activity on the identification of the planetary albedo, in the context of the observing conditions imposed by the new joint ESA-Switzerland optical photometric space mission CHEOPS (CHaracterizing ExOPlanet Satellite). The
main objective of CHEOPS is determining with a very high precision the radii of planets smaller than Saturn that are known to orbit a selected number of bright stars \citep{Fortier}. In this way, it will be possible to determine the composition of super-Earths and impose constraints on planetary formation. Moreover, the mission will address several other problems. One of them is the detection of phase curve modulations induced by exoplanets on the stellar flux and thus to estimate the albedo and deduce information about the atmosphere. 
   
   We have simulated synthetic stellar light curves that include a modulation caused by the presence of spots, a planetary phase curve, and an additional instrumental noise term. We modeled the instrumental noise as white noise, with a level comparable to the predicted one for CHEOPS\footnotemark . We then estimated the albedo of the simulated planet, assuming that we knew all its other properties, such as the orbital period and the radius. 
   
   We decided not to include the beaming effect and the ellipsoidal modulation for two reasons. The first is that for all the planets considered, these two effects are generally negligible, except for the case of giant planets with short semimajor axes \citep[e.g.,][]{Angerhausen}. Second, all the planet parameters on which they depend can be determined through other methods and are known for the CHEOPS targets. For example, the beaming effect depends on the mass, which is directly determined from the radial velocity semi-amplitude. In turn, the ellipsoidal modulation depends on planet and stellar parameters that again can be deduced from spectroscopic observations of the system \citep{Lillo-Box}.  For these reasons, as we show in section \ref{Extra}, the addition of such components does not have a significant effect on the precision with which the albedo is derived by our model. 
   
   We also decided not to model the primary and the secondary transit. Without adding the two eclipses in our model, we aim at revealing non-transiting planets through their phase light curves alone, as shown in \citep{Crossfield}. Many of the planets detected through the radial velocity technique do not transit their host because of the orientation of their orbits. We can detect planetary atmospheres of transiting planets through transmission \citep{Deming, Kirk, Kreidberg, Pont} and emission spectroscopy \citep{Charbonneau2, Knutson2, Knutson1}. In the case of a non-transiting planet, these techniques cannot be applied, but detecting the phase curve still allows for a study of its atmosphere.
   
   With these assumptions, we performed an analysis of the possible change in detection precision of the albedo when we account for the uncertainty of the planetary radius and semi-major axis, as determined through the primary transit. 
   
   In Sect. \ref{simulations_model} we present the model adopted to build the stellar simulations and the choices regarding the parameters of the star and of the planet. We also explain how we simulated the instrumental noise. In Sect. \ref{GPMCMC} we describe the method adopted to analyze the simulated light curves, and in the following paragraph, we describe the blind tests that proved the reliability of this method. In Section \ref{results} the results of the data analysis are reported. In Sect. \ref{Extra} we assess what happens when the beaming effect and the ellipsoidal modulation are added to the model and when additional priors are inserted around the planetary radius and the semi-major axis. Sect. \ref{holes} shows an additional test for determining the albedo of the exoplanets in presence of gaps in the observations,
such as are predicted for CHEOPS. In Sect. \ref{Kepler_star_sec} we explain the limitations in the application of our model to a real planet by performing a test on Kepler-7b. We additionally report a test in which we added  a simulated planetary light curve modulation to a periodic Kepler star. The last section presents the conclusions.
   
   \footnotetext{It could have been more realistic if instead of white noise we had added a realistic red noise to our simulation. However, it is not obvious or clear so far what the real CHEOPS noise will be.}
   
\section{Synthetic light curves}
\label{simulations_model}
\begin{figure*}
\centering
\includegraphics[width=20cm]{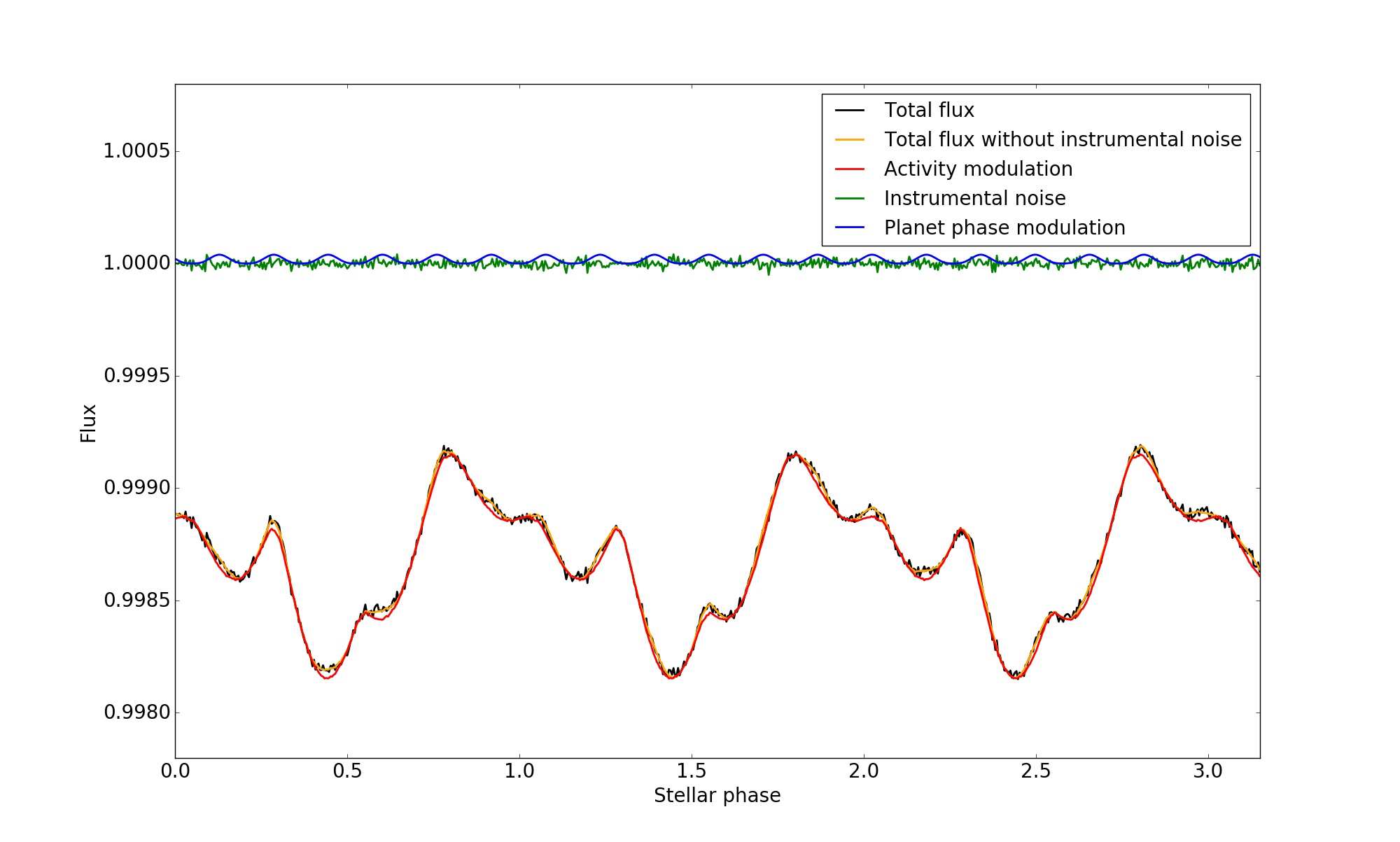}
\caption{Typical phase light curve. It shows the normalized stellar flux as a function of the stellar phase. The green and blue lines represent the instrumental noise and the planet phase modulation, respectively, both shifted by $1$. The planet phase modulation is built with an albedo of $0.3$, a planetary radius $R_p = 0.1 \; R*$ , and properties as listed in Table \ref{tab7}. The red line is the stellar activity modulation as listed in Table \ref{tab6}, the black line shows the total flux, and the orange line is the total flux without the instrumental noise. This example light curve is also adopted for most of the tests performed in this paper.}
\label{plots_light_curve}
\end{figure*}
As a first step of the work, we have simulated stellar light curves as a function of the stellar phase, $\Phi$. Each light curve includes three different components:
\begin{itemize}
\item the reflected light phase modulation due to the planet
\item the stellar activity modulation 
\item a white noise at the same level of the instrumental noise.
\end{itemize}
Previous works have used the secondary transit and the planetary phase curve at the same time to determine the albedo of the exoplanet \citep[e.g.,][]{Esteves, Angerhausen}. In certain cases, the secondary transit cannot be directly identified because of the stellar activity modulation. An example is planet Kepler-91b, for which \citet{Lillo-Box} have fitted the primary transit and have analyzed the remaining phase curve taking into account the reflected light component, the ellipsoidal modulation, and the beaming effect. They finally predicted the possible position of the secondary transit, which seemed to be contaminated by an additional activity modulation. Because of this higher uncertainty in identifying the secondary transit, we chose not to include it in the model and verified whether it might be possible to identify the albedo after fully characterizing the planetary orbit. This also suggests the possibility of a future development of our tool in the case of non-transiting planets. For example, \citet{Crossfield} have been able to determine the presence of an additional non-transiting planet with the planetary phase curve analysis alone. 

        The CHEOPS timing will be $\text{one}$ minute, but its observations are predicted to have several gaps. In particular, there are going to be two types of gaps. The first gap occurs because the satellite crosses the South Atlantic Anomaly. Here, the terrestrial magnetic field generates loops that trap high-energy particle. When many of these particles impact the detector of the satellite, they produce glitches in the observations, which are thus too noisy and are removed from the data. These interruptions only depend on the position of the satellite above the Earth and are independent of the selected target. The other gaps depend on the observed star and occur because the target is occulted or too close to the Earth limb, the Sun, or the Moon. All these interruptions in the observations will be several minutes long each, but all together are expected to cover a significant fraction of the data. To have a much longer sampling than the predicted gaps, we chose to simulate curves with$ \; \text{a}$ timing of
two hours, which means that we will have 12 points per day. In this way, the results of our test are expected to remain relatively
unchanged with respect to the more realistic curves with gaps, because the gaps are shorter than the bin size and their overall impact is only to decrease the signal-to-noise ratio of each bin. We show this in Section \ref{holes}.

\subsection{Reflected light modulation}
\label{simulation_planet}
        We expressed the planet orbital phase modulation using the general equation reported in \citet{Madhusudhan} for the ratio between the planet flux $F_P$ and the stellar flux $F_*$:
\begin{equation}
\frac{F_P}{F_*} = A_P \; f(z)
,\end{equation}
where $f(z)$ is the planetary phase function, $z$ depends on the orbital inclination $i$ and on the the orbital phase, and $A_P$ is the associated amplitude. The choice of how to model the planetary phase function depends on the hypothesis we impose on the atmosphere of the planet. \citet{Madhusudhan} performed a comprehensive study on the different models for the albedo, including the model associated with Rayleigh scattering from the atmosphere. In the present work, we have decided to adopt the most simple model, as was done in \citet{Esteves} and \citet{Angerhausen}). It consists of describing the planet as a Lambertian sphere \citep{Russell}, which means that we assume the planet to be a perfect sphere with an atmosphere reflecting the stellar flux
isotropically. In this case, the planetary phase function is given by \citet{Angerhausen}
\begin{equation}
f(z) = \frac{\sin (z) + (\pi - z) \cos(z)}{\pi}
.\end{equation}
        
        The amplitude of the phase function is
\begin{equation}
\label{amplitude}
A_P = A_g \left(\frac{R_P}{r}\right)^2
.\end{equation}
$A_g$ is the geometric albedo, $R_P$ is the radius of the planet, and $r$ is the planet-star distance \citep{Lillo-Box}. Since from now on we simulate circular orbits, $r$ is equal to the semi-major axis $a$, so that $R_p/r$ becomes $R_p/a$.

\subsection{Stellar activity}
As stellar activity, we typically indicate a group of phenomena that occur on the stellar surface \citep{Berdyugina1}. Of the phenomena that can influence the photometric stellar signals, the most evident are spots and plages, generated in the areas from which the magnetic flux tubes emerge or enter the stellar surface \citep{Kitiashvili}. The spots are characterized by a lower temperature than the average stellar surface temperature and they therefore look darker than the rest of the stellar surface. Since spots generate a decrement in the stellar light curve, not taking into account their presence in photometric observations determines a bias in the characterization of the planet \citep[e.g.,][]{Oshagh2, Barros}. Plages are generally considered as the hotter counterpart of spots and generate opposite effects on the stellar light curves. Their effect is less significant because they are characterized by a lower temperature contrast than spots \citep[e.g.,][]{Berdyugina1, Meunier}. For this reason, we decided to model stellar light curves in presence of spots alone. 

        For the transits, the spot features represent noise. Unless they are very large, they can affect the characterization of a planet, but they do not hide it completely \citep{Barros_1}. Instead, in case of a planetary phase light curve, the stellar activity can be a much stronger signal and it can hide the planetary phase curve. Even for quiet stars, the signal of the phase curves can be much lower than the stellar activity. For instance, the planetary phase modulation is usually lower than $200 \; ppm$ \citep[e.g.,][]{Angerhausen, Lillo-Box} and depends on the planet radius, while the stellar activity can reach much higher levels, even $10^4 \; ppm$, as in the case of the star Corot-7 \citep{Leger}. 

        To model the stellar activity, we adopted the SOAP-T tool \citep{Oshagh1}. This program uses a numerical method to simulate the stellar light curve of a spotted star, orbited by one planet, and generates the radial velocity pattern of the star from the flux modulation.
        
        To avoid introducing the transit feature in the light curve, we did not simulate the planet within SOAP-T. This means that the only planetary component introduced in the light curve comes from the phase light curve modeled as in Sect. \ref{simulation_planet}.
                
\subsection{Instrumental noise}
To render the light curve more realistic and more similar to CHEOPS observations, we introduced an additional component that simulates the instrumental noise.

        To simplify the production of the synthetic simulations, we modeled the noise component as Gaussian noise with a standard deviation comparable to the value achieved by the satellite CHEOPS for different magnitudes of the star. In Table \ref{tab1} we report the values of CHEOPS noises for the stellar magnitudes we adopted.
\begin{table}
\caption{CHEOPS standard deviations for stars of different magnitudes. Courtesy of the CHEOPS consortium.}             
\label{tab1}      
\centering                          
\begin{tabular}{c c}        
\hline\hline                 
 Magnitude & Noise for two-hour timing \\    
\hline                        
   $6.5$    & $14 ppm$  \\      
   $8$      & $17 ppm$ \\
   $10$     & $29 ppm$ \\
\hline                                   
\end{tabular}
\end{table}

\subsection{Final production of the light curves}
\label{simulations}
We built the light curves to be tested by summing the three components described in the previous sections:
\begin{equation}
\frac{F_{total}}{F_*} = \frac{F_p}{F_*} + \frac{F_{*,spotted}}{F_*} + \frac{F_{noise}}{F_*}
,\end{equation}
where the first term is the planet phase light curve modulation, the second term is the light curve produced by SOAP-T, and the third term is the noise. The total obtained flux was normalized with respect to the stellar flux and thus has no units. In Figure \ref{plots_light_curve} we show a typical stellar light curve. In particular, we show the comparison between the instrumental noise, the planetary phase light curves, and then the stellar activity pattern, the total flux, and the total flux without instrumental noise as a function of the stellar phase. The spots are spread throughout the light curve, and the amplitude of the activity signal is much higher than the reflected light component of the planet, preventing visual identification.

\section{Data analysis method}
\label{GPMCMC}
With the simulated light curves at hand, we adopted a Markov
chain Monte Carlo (MCMC) parameter fitting analysis to establish whether we were able to recover the planetary albedo in presence of the stellar activity and the instrumental noise. We assumed that all the planetary parameters were known (e.g., from a transit) and that the only parameter to be determined was the albedo itself. To do this, we needed to model the stellar activity signal and distinguish it from the reflected light component of the planet.

\subsection{Gaussian process for modeling the stellar activity}
 \begin{figure*}
        {\includegraphics[width=18cm]{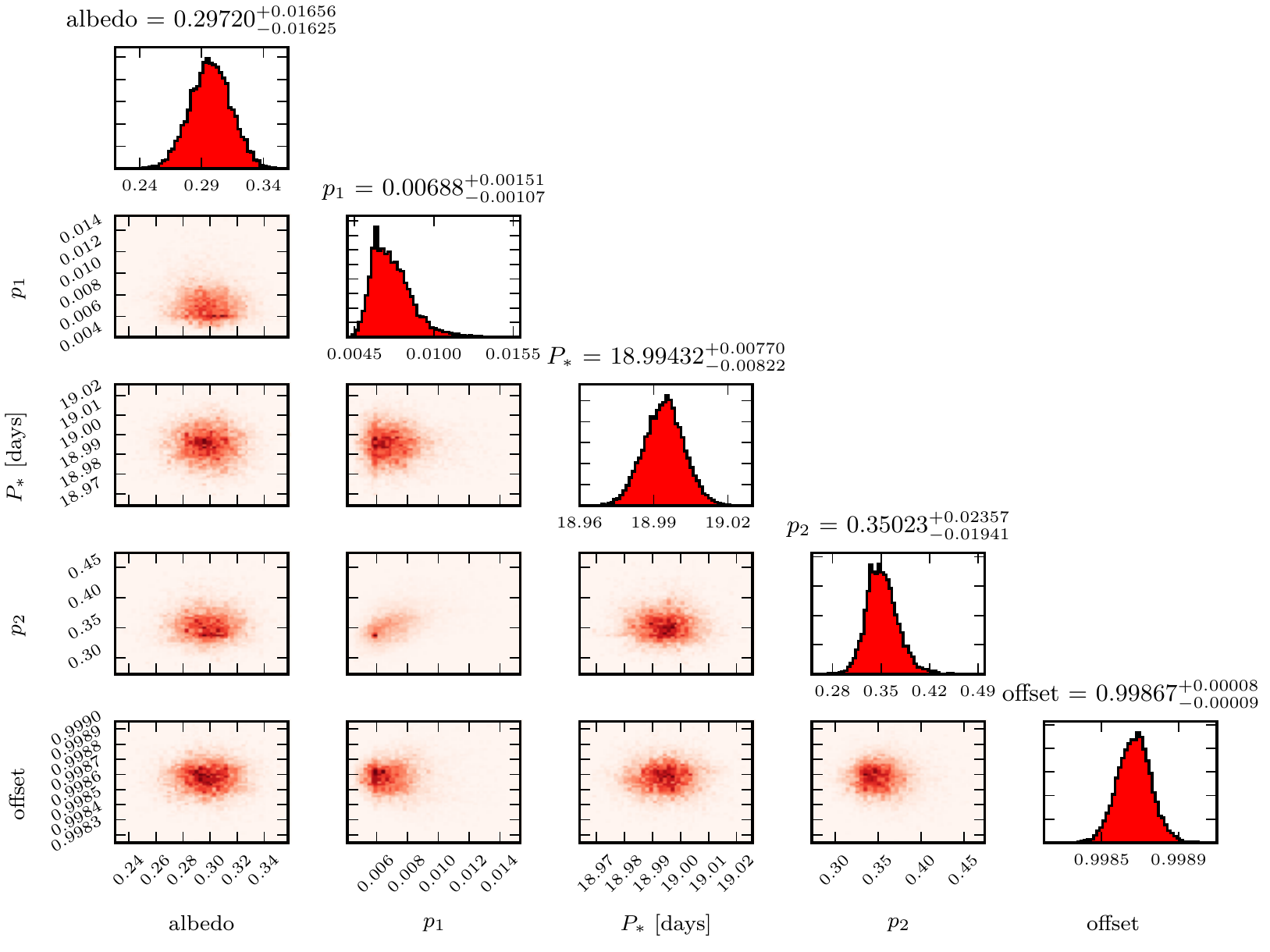}}
         \caption{1D and 2D posterior distributions for the parameters for a star rotating with a period of $19 \; \text{days,}\text{}$  with an orbiting planet with radius $0.1 \; R_*$ , observed for $13$ full orbital periods, in presence of the four-spot activity pattern in Table \ref{tab6}. The input albedo is $0.3$.}
         \label{Corner_good}
\end{figure*}
\begin{table}
\caption{Adopted priors for the five parameters of the MCMC; $P_{0,*}$ represents the original value of the stellar rotation used to build the simulation, $F_{mean}$ is the flux average, and $ptp$ is the peak-to-peak variation of the light curve.}         
\centering                         
\begin{tabular}{c c c}       
\hline\hline                
 Parameter & Prior & Interval \\    
\hline                        
   $A_g$    & Uniform     & $\left[0;1\right]$ \\     
   $P_*$    & Gaussian    & $N(P_{0,*},3)$\\
   $p_1$    & Log-Uniform & $\left[10^{-4};600\right]$\\
   $p_2$    & Log-Uniform & $\left[5\times 10^{-5};2 \times 10^{4}\right]$\\
   Offset   & Uniform     & $\left[F_{mean} - 2\;ptp; F_{mean}+2\;ptp\right]$\\ 
\hline                                  
\end{tabular}
\label{tab2}   

\end{table}
The spot variability is difficult to model analytically  because the parameters that influence the shape of the features are highly degenerate. For example, the depth of an activity modulation depends on both the temperature contrast and the size of the spot the modulation originates in, but it might likewise be the
case that the same feature is due to the overlapping of two different spots. This significantly increases the number of possible solutions for modeling the same stellar light curve. Because we cannot
directly observe the spot distribution, we need to adopt a different strategy.

        Recent works on planetary detection in presence of stellar activity have adopted Gaussian processes (GP) for modeling the spot modulation, both in radial velocity and photometry \citep{Faria, Haywood, Rajpaul}. This method consists of treating the stellar activity noise as a correlated noise, since the spots move on the stellar surface through the rotation of the star. A GP is in part defined by its covariance function. For activity-induced signals, a quasi-periodic covariance function is the most common choice \citep{Rajpaul}.
        
        We here decided not to include the aperiodic component because we did not simulate spot evolution, which is not predicted to be observed within short observations \citep[for the lifetime of spots, see ][]{Berdyugina1}. Consequently, our covariance is fully periodic and is given by
\begin{equation}
\Sigma_{ij} = p_1 exp\left[-\frac{2\sin^2\frac{\pi (t_i-t_j)}{P_*}}{p_2^2}\right] + \sigma^2 \delta_{ij}
.\end{equation}
        Here, $p_1$ is the amplitude of the correlations. Then, the exponent represents the periodic correlation. It shows that the stellar noise varies periodically with the stellar rotation and decays on a timescale of $p_2$. The last part of this covariance function includes a diagonal component that depends on the instrumental noise $\sigma$ ($\delta _{ij}$ is the Kronecker delta). From now on, $p_1$, $p_2$ and $P_*$ are also called hyper-parameters of the GP.

        This choice of covariance function is in line with our simulated light curves, which do not contain spot evolution (and are therefore completely periodic by construction), but it might be less adequate for the analysis of more general light curves. Furthermore, we assumed that the level of instrumental noise $\sigma$ is known and fixed to the estimated CHEOPS noise for a given stellar magnitude. Unlike other studies, which applied GP on real data, we decided not to introduce a jitter parameter for the tests on the synthetic light curves. We tried, indeed, to add it in the beginning, but it was always very close to zero, which showed it was meaningless to introduce it as an additional parameter in these special cases. We used the \texttt{george} package to perform the GP regression \citep{Ambikasaran}.

\subsection{Analysis method}
To sample from the posterior distributions for the parameters of our model, we used the tool \texttt{emcee} \citep{Foreman} to perform an MCMC. We have five parameters: the geometric albedo $A_g$, the stellar rotation $P_*$, the amplitude of the correlation $p_1$, the timescale decay of the periodic modulation $p_2$ ,
and an offset to fit the average value of the light curve. 

        In Table \ref{tab2} we report the selected priors for each parameter. Since the albedo can adopt any value between $0$ and $1$ and is otherwise unconstrained a priori, we assumed a uniform prior for it. The geometric albedo is rarely higher than $1$, but this is not the case for the exoplanets we model here. The stellar rotation $P_*$ is difficult to estimate. We here assumed that we consider stars whose rotational period has
been reported in previous papers, with a typical uncertainty of$ \; \text{three days.}$  Consequently, the stellar rotation prior is a Gaussian, centered on the initial value of $P_*$ and with a standard deviation of three$ \; \text{days}$. The GP coupled with an MCMC allowed us to derive the stellar rotation with much better precision than in other methods, such as a Lomb-Scargle periodogram or an autocorrelation function technique, as has been proved by \citet{Angus}. If the rotational period is not known, the data can first be analyzed with these other methods to determine the prior on $P_*$ to be adopted for our analysis. For the other hyper-parameters, $p_1$ and $p_2$, we chose very wide log-uniform priors, with the intervals listed in Table \ref{tab2}. The last parameter of the MCMC is the offset of the GP, for which we used a uniform prior. This prior is centered on the average flux level $F_{mean}$ with a width of four times the peak-to-peak variation of the light curve.
        
        We  adopted the reflected planetary phase curve model of Section \ref{simulation_planet} as the planetary model and used the GP to model the stellar activity. The overall likelihood is expressed as a multivariate Gaussian distribution \citep{Ambikasaran, Faria}. After choosing the priors, the model, and the likelihood, we randomly extracted the initial parameter values of 30 MCMC chains from the prior distribution. For each simulation, we ran the MCMC with a 500-step burn-in and then again sampled the chains for 1000 steps. In general, we obtained 30000 effective samples from the posterior distribution function for each simulation.

        Then, we chose to determine the best fit by estimating the median values of the posterior distributions for each parameter, since the posteriors were Gaussians. When the MCMC was unable
to fit the planet curve, the recovered albedo was $0$, therefore its posterior was hyperbolic with a peak on $0$. In these cases, which were identified with a visual check, we adopted the mode as best-fit parameter. The $1 \sigma$ uncertainties were determined as differences between the best-fit value and the $16^{th}$ and $84^{th}$ percentiles, respectively.
        
        In Figure \ref{Corner_good} we report an example of the
output for the parameter posterior distributions for a rotating
star with period of  $19$ days, with an orbiting planet with
  a radius $0.1 \; R_*$ that is observed for 13 full orbital periods.

\section{Reliability test}
\begin{table}
\caption{Stellar and planetary properties common for all the performed blind tests.}            
\label{tab3}      
\centering                          
\begin{tabular}{c c c}        
\hline\hline                                      
   Stellar radius                          & $R_*$ & $1 \; R_{\odot}$ \\      
   Stellar inclination                     & $I$   & $90^{\circ}$\\
   Stellar temperature                     & $T_*$ & $5778$ K   \\
   Linear limb-darkening coefficient       & $c_1$ & $0.29$\\
   Quadratic limb-darkening coefficient    & $c_2$ & $0.34$\\
   Planet radius                           & $R_p$ & $0.1\; R_*$ \\
   Time of mid-transit                     & $t_0$ & $0.3$ days \\      
   Eccentricity                            & $e$   & $0$\\
   Argument of periastron                  & $w$ & $0^{\circ}$   \\
   Inclination of the orbital plane        & $i$ & $89^{\circ}$\\
   Projected spin-orbit misalignment angle & $\lambda$ & $0^{\circ}$\\    
\hline                                   
\end{tabular}
\end{table}

\begin{table}
\caption{Spot properties used to generate the activity patterns of the blind tests with SOAP \citep{Oshagh2}. The pattern labeled $a$ has been adopted for tests 1-3, the $b$ pattern for tests 4-6, and the $c$ pattern for the last three tests. This information was unknown by the person that performed the analysis.}             
\label{tab5}      
\centering                          
\begin{tabular}{c c c c c}        
\hline\hline 
\\                
 Pattern & Spot & Longitude   & Brightness & Size ($R_*$) \\    
\hline                        
         & 1   & $0   ^{\circ}$ &  $0.50$     & $0.080$ \\  
         & 2   & $55  ^{\circ}$ &  $0.48$     & $0.075$ \\
  $a$    & 3   & $120 ^{\circ}$ &  $0.52$     & $0.081$ \\  
         & 4   & $174 ^{\circ}$ &  $0.48$     & $0.079$ \\  
         & 5   & $227 ^{\circ}$ &  $0.50$     & $0.083$ \\  
         & 6   & $290 ^{\circ}$ &  $0.49$     & $0.076$ \\  
\hline
         & 1   & $0   ^{\circ}$ &  $0.50$     & $0.040$ \\  
  $b$    & 2   & $20  ^{\circ}$ &  $0.48$     & $0.045$ \\
         & 3   & $35  ^{\circ}$ &  $0.52$     & $0.041$ \\  
         & 4   & $121 ^{\circ}$ &  $0.48$     & $0.049$ \\  
\hline
         & 1   & $0   ^{\circ}$ &  $0.50$     & $0.02$ \\  
  $c$    & 2   & $34  ^{\circ}$ &  $0.48$     & $0.025$ \\
    
\hline                                   
\end{tabular}
\end{table}
 \begin{figure}
        {\includegraphics[width=9.3cm]{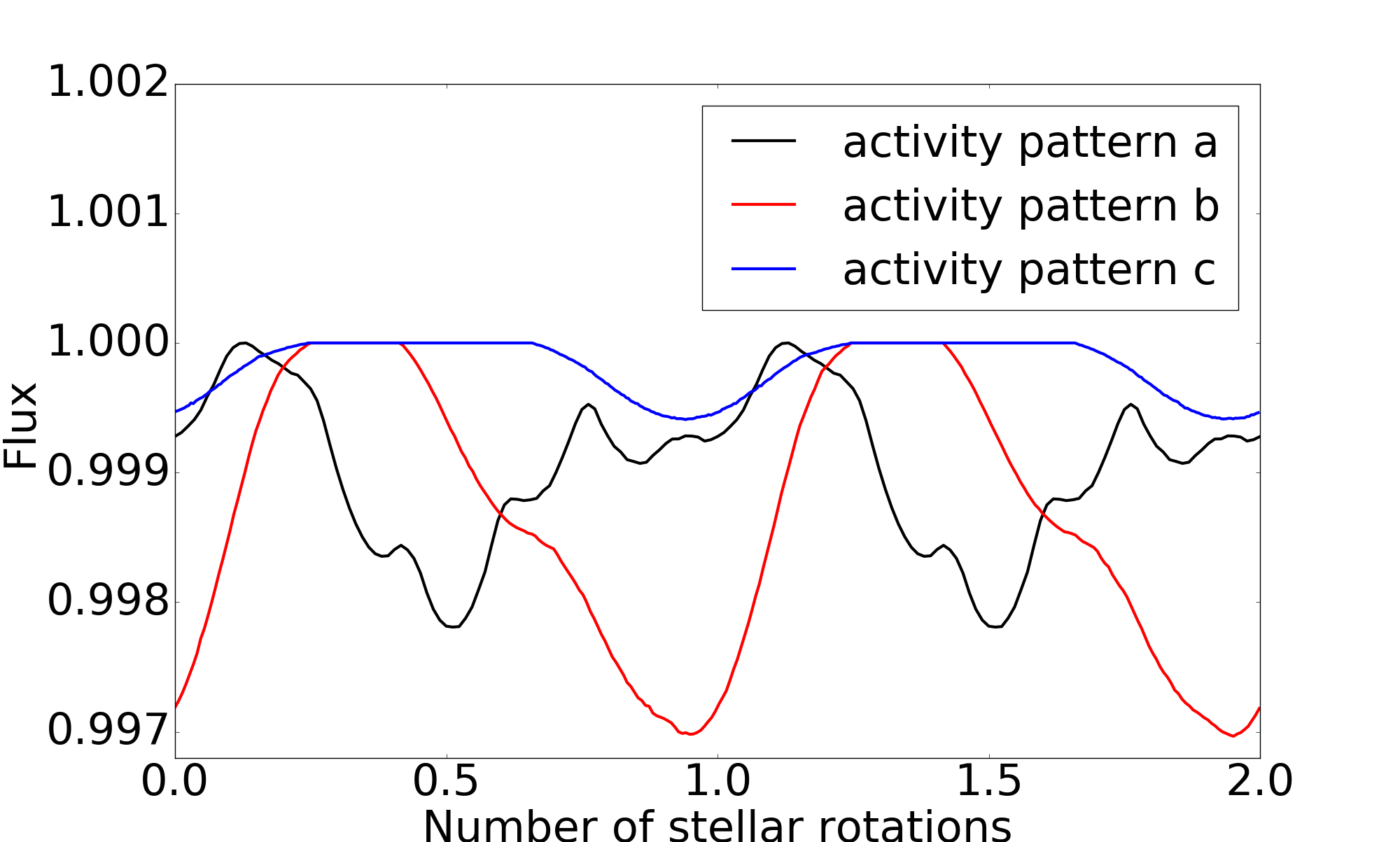}}
         \caption{Comparison between patterns $a$, $b,$ and $c$ adopted in the blind tests. Their properties are reported in Table \ref{tab5}.}
         \label{activity_patterns}
\end{figure}
\begin{table*}
\caption{Input properties and recovered parameters for the blind tests.}             
\label{tab8}      
\centering                          
\begin{tabular}{c | c c c c c | c c c c}        
\hline\hline

   \# & \multicolumn{5}{c|}{Input properties} & \multicolumn{4}{|c}{Recovered parameters} \\
\hline 
      & Duration & Pattern & $P$      & $A_g$ & $P_*$    & $A_g$ & $P_*$    & $p_1$ & $p_2$  \\
      & $(days)$ &         & $(days)$ &       & $(days)$ &       & $(days)$ &       &        \\
\hline

               & \multicolumn{5}{c|}{} & \multicolumn{4}{|c}{} \\
   1           & $10.66$ & a & $2.2$ & $0.8$ & $7.00$ & $0.828^{\left[+0.037\right]}_{\left[-0.035\right]}$ & $7.000^{\left[+0.003\right]}_{\left[-0.003\right]}$  & $0.0049^{\left[+0.0009\right]}_{\left[-0.0006\right]}$  & $0.319^{\left[+0.018\right]}_{\left[-0.016\right]}$ \\
               & \multicolumn{5}{c|}{} & \multicolumn{4}{|c}{} \\
   2           & $10.25$ & a & $1.5$ & $0.2$ & $7.00$ & $0.212^{\left[+0.018\right]}_{\left[-0.017\right]}$ & $7.000^{\left[+0.003\right]}_{\left[-0.003\right]}$  & $0.0048^{\left[+0.0009\right]}_{\left[-0.0007\right]}$  & $0.302^{\left[+0.016\right]}_{\left[-0.017\right]}$ \\
               & \multicolumn{5}{c|}{} & \multicolumn{4}{|c}{} \\
   3           & $10.25$ & a & $1.2$ & $0.1$ & $7.00$ & $0.129^{\left[+0.020\right]}_{\left[-0.019\right]}$ & $7.000^{\left[+0.003\right]}_{\left[-0.003\right]}$  & $0.0049^{\left[+0.0009\right]}_{\left[-0.0007\right]}$  & $0.301^{\left[+0.017\right]}_{\left[-0.017\right]}$ \\
               & \multicolumn{5}{c|}{} & \multicolumn{4}{|c}{} \\
   4           & $17.08$ & b & $2.2$ & $0.3$ & $12.30$ & $0.319^{\left[+0.012\right]}_{\left[-0.012\right]}$ & $12.299^{\left[+0.003\right]}_{\left[-0.003\right]}$ & $0.0091^{\left[+0.0028\right]}_{\left[-0.0021\right]}$  & $0.729^{\left[+0.109\right]}_{\left[-0.105\right]}$ \\
               & \multicolumn{5}{c|}{} & \multicolumn{4}{|c}{} \\
   5               & $18.75$ & b & $1.5$ & $0.5$ & $12.30$ & $0.469^{\left[+0.013\right]}_{\left[-0.011\right]}$ & $12.295^{\left[+0.005\right]}_{\left[-0.004\right]}$ & $0.0092^{\left[+0.0030\right]}_{\left[-0.0020\right]}$  & $0.790^{\left[+0.361\right]}_{\left[-0.327\right]}$ \\
               & \multicolumn{5}{c|}{} & \multicolumn{4}{|c}{} \\
   6               & $20.42$ & b & $1.2$ & $0.4$ & $12.30$ & $0.400^{\left[+0.006\right]}_{\left[-0.006\right]}$ & $12.296^{\left[+0.005\right]}_{\left[-0.005\right]}$ & $0.0085^{\left[+0.0023\right]}_{\left[-0.0016\right]}$  & $0.616^{\left[+0.041\right]}_{\left[-0.037\right]}$ \\
               & \multicolumn{5}{c|}{} & \multicolumn{4}{|c}{} \\
   7               & $19.41$ & c & $2.2$ & $0.6$ & $19.74$ & $0.609^{\left[+0.012\right]}_{\left[-0.012\right]}$ & $20.780^{\left[+0.965\right]}_{\left[-0.777\right]}$ & $0.0029^{\left[+0.0015\right]}_{\left[-0.0011\right]}$  & $1.134^{\left[+0.341\right]}_{\left[-0.212\right]}$ \\
               & \multicolumn{5}{c|}{} & \multicolumn{4}{|c}{} \\
   8               & $26.08$ & c & $2.7$ & $0.35$ & $19.74$ & $0.340^{\left[+0.015\right]}_{\left[-0.015\right]}$ & $19.739^{\left[+0.028\right]}_{\left[-0.030\right]}$ & $0.0041^{\left[+0.0024\right]}_{\left[-0.0018\right]}$  & $1.511^{\left[+0.319\right]}_{\left[-0.250\right]}$ \\
               & \multicolumn{5}{c|}{} & \multicolumn{4}{|c}{} \\
   9           & $24.42$ & c & $1.2$ & $0.15$ & $19.74$ & $0.155^{\left[+0.005\right]}_{\left[-0.005\right]}$ & $19.732^{\left[+0.022\right]}_{\left[-0.025\right]}$ & $0.0044^{\left[+0.0027\right]}_{\left[-0.0025\right]}$  & $1.601^{\left[+0.469\right]}_{\left[-0.347\right]}$ \\
               & \multicolumn{5}{c|}{} & \multicolumn{4}{|c}{} \\
\hline                                   
\end{tabular}
\end{table*}

        To verify whether our analysis method and code worked as desired, we first performed a series of blind tests. In these, one of the co-authors prepared nine mock light curves and passed us all the planet parameters, except for the albedo, and a range of possible rotational period values. The goal was to check whether we were able to derive the correct albedo and stellar rotational period without further information. In Table \ref{tab3} we report the initial parameters, which were common for all the blind tests. The star had the radius and temperature of the Sun, an inclination $I$ of the rotational axis perpendicular to the line of sight of the observer, and the stellar equator was seen edge-on. The last two quantities are the limb-darkening coefficients \citep{Claret}, which were chosen within the Kepler bandpass tables in \citet{Claret1}. The modeled planet was a Jupiter-sized planet, with varying period and albedo. 

        Table \ref{tab5} shows the properties of the spots for each of the three adopted stellar activity patterns, which were not known by the person that applied the MCMC on the simulations. The patterns are also displayed in Figure \ref{activity_patterns}. Tests 1, 2 and 3 present the most active pattern, while the least active was assigned to the last three tests. For each test, we
report in Table \ref{tab8} the remaining input parameters and the outputs of the analysis. The second column shows the length of the observation and the third column the adopted activity pattern. The fourth column shows the period of the simulated planet, expressed in days. This information was known before the analysis was performed. The fifth and sixth columns show the the input albedo and stellar rotation, which were unknown by the person that applied the MCMC to the simulations. For $P_*$ the only information that was passed to the analyzer was an interval around the input value, not exactly centered on it. The remaining four columns of Table \ref{tab8} report the output of the MCMC analysis with the albedo, the recovered $P_*$, and the hyper-parameters $p_1$ and $p_2$. We did not report the offset because it was almost the same for all tests, $0.999$. 
        
        In all the tests, we obtained albedo values that were
compatible with the original values within $1 \sigma$ . The stellar rotations were also recovered with high precision, within $2 \sigma$. The only exception is the seventh case, where the rotational period was overestimated and the error was about one$ \; \text{day}$. For this test, the observational length was very close to the rotational period. The reason behind this is better explained in the next section, which shows that we need more than one stellar rotation to derive reliable results for both $P_*$ and $A_g$. For the tests with the same activity pattern, the hyper-parameters and the offset were very similar between each other, showing that the GP regression works properly. 

        These tests demonstrate that our analysis can distinguish the reflected light component of the planet from the stellar noise for a wide range of planetary and stellar properties.

\section{Results}
\label{results}
        To investigate the limits of the method in recovering the stellar period of rotation and the albedo, we applied our analysis tool on a series of$ \; \text{}$ 60-day-long simulations with different $P_*$. Depending to what we wished to investigate, we varied the observation length, the planetary radius, the albedo, the orbital period, and the spot dimension.

\subsection{Simulation properties }

        We decided to model the case of a $6.5$ magnitude Sun-like star, whose properties are those used for the blind tests (see Table \ref{tab3}). We simulated a stellar activity pattern with four spots placed at different longitudes in order to cover the full stellar light curve and reproduce a realistic spot modulation. In Table \ref{tab6} we report the selected properties for the spots. The latitudes were fixed to $0$, so that the spots appear at the stellar equator. The spot sizes were fixed as well. 
        For the planet, we decided to fix the physical quantities reported in Table \ref{tab7}.

\begin{table}
\caption{Spot properties introduced in SOAP-T \citep{Oshagh1}.}             
\centering                          
\begin{tabular}{c c c c}        
\hline\hline  
\\               
 Spot & Longitude      & $\Delta T$ ($K$) & Size ($R_*$) \\    
\hline                        
 
1     & $270 ^{\circ}$ & $400$  & $0.045$\\
2     & $80 ^{\circ}$  & $500$  & $0.045$\\
3     & $250 ^{\circ}$ & $663$  & $0.045$\\
4     & $340 ^{\circ}$ & $700$  & $0.045$\\
\hline
\end{tabular}
\label{tab6}      
\end{table}

\begin{table}
\caption{Fixed planetary properties.}             
\centering                          
\begin{tabular}{c c c}        
\hline\hline    
\vspace{0.1pt}             
   Time of mid-transit                     & $t_0$ & $0.2$ days \\      
   Eccentricity                            & $e$   & $0$\\
   Argument of periastron                  & $w$   & $0^{\circ}$   \\
   Inclination of the orbital plane        & $i$   & $90^{\circ}$\\
   Projected spin-orbit misalignment angle & $\lambda$   & $0^{\circ}$\\ 
\hline                                   
\end{tabular}
\label{tab7}
\end{table}
\subsection{Lower limit for the observation length}
        As a first step, we fixed the planetary radius to $R_p = 0.1 \; R_*$ and the orbital period to $P = 3 \; \text{days}$. Then, we produced simulations with increasing durations, starting with a simulation that covered an entire orbital period, which
we increased by$ \; \text{3 days}$ each time. The longest simulation covered $60 \; \text{days}$, that is, $20$ entire orbital periods. In this way, we aimed at exploring the minimum length of observations necessary for detecting the planetary albedo together with the stellar rotation. 
        
\begin{figure*}
\centering
\includegraphics[width=9.1cm]{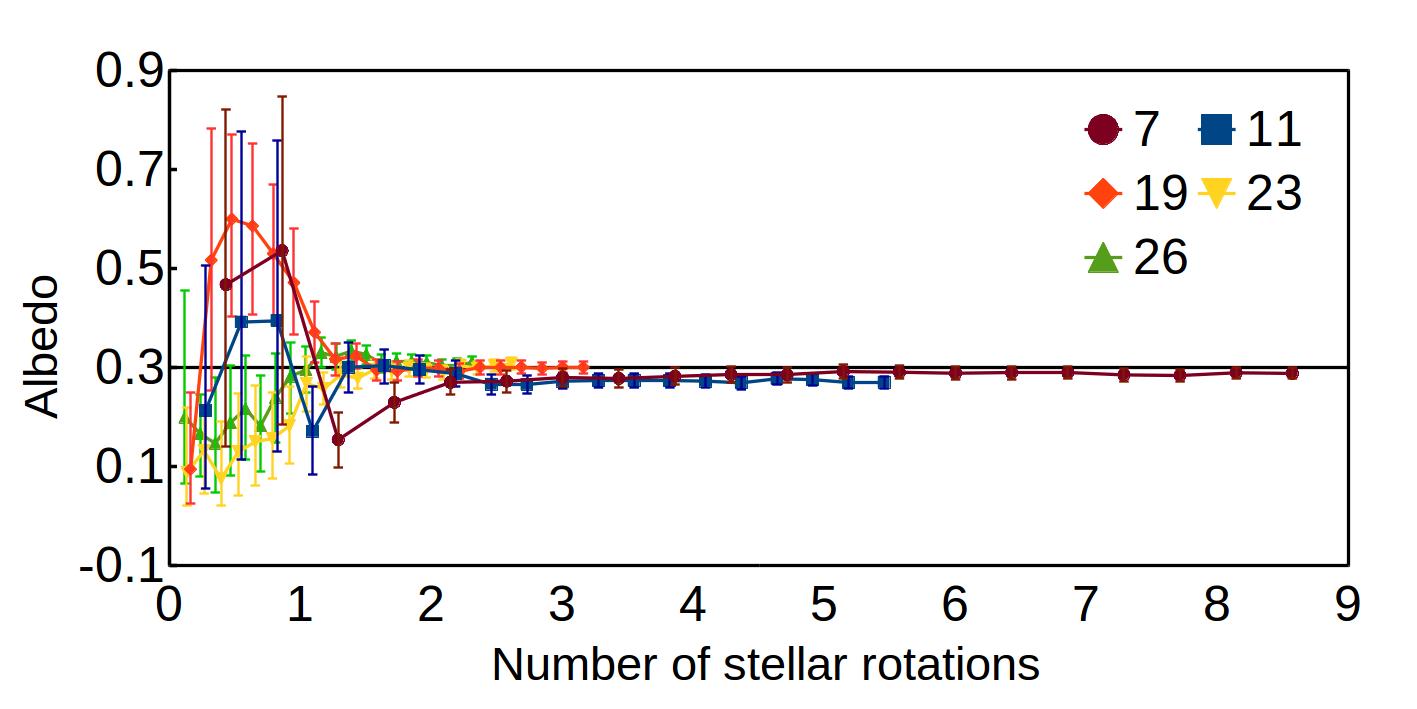}
\includegraphics[width=9.1cm]{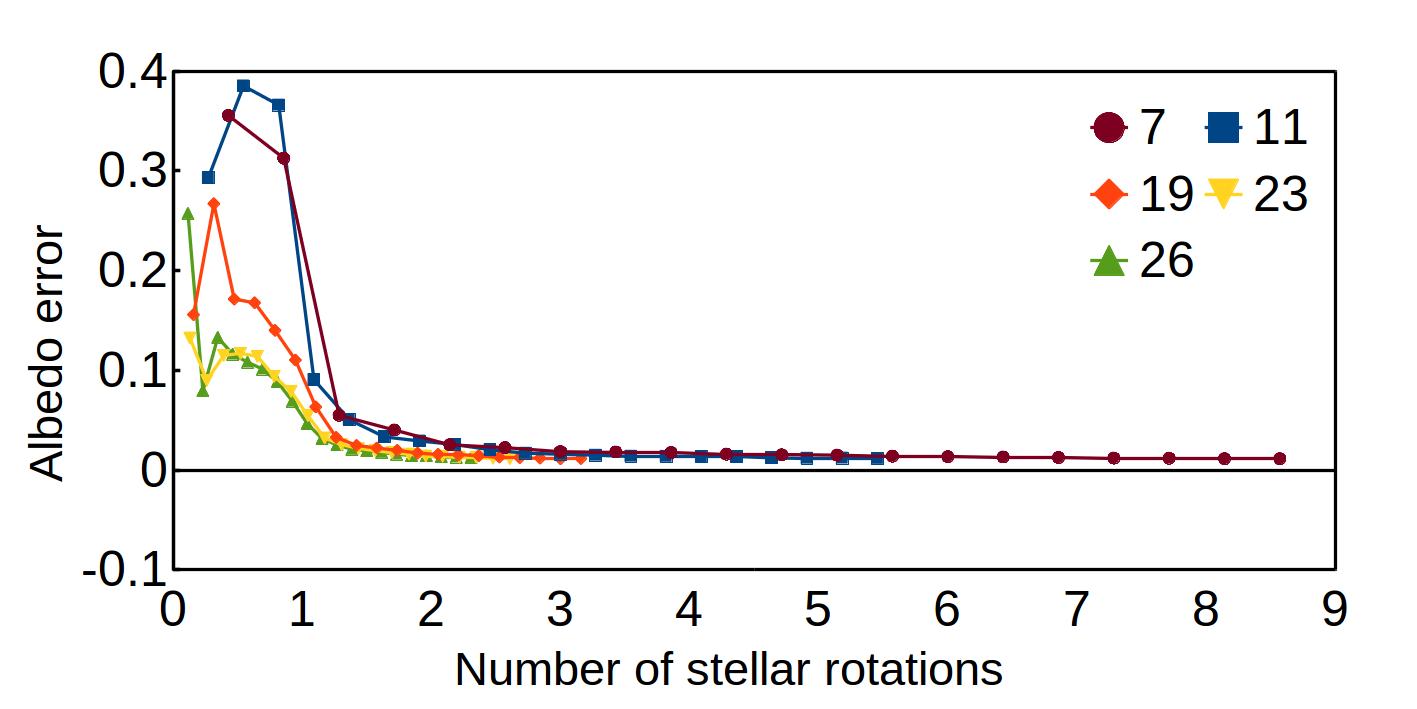}
\caption{Plots of the albedo and relative errors for the simulations obtained with $P_* = 7$, $11$, $19$, $23,$ and $26 \; \text{days}$ and increasing observational lengths. The input stellar properties are reported in Table \ref{tab3}, while the planetary properties are listed in Table \ref{tab7}. The activity pattern is the one of Table \ref{tab6}. The initial albedo is $0.3$. In the left panel, we report the albedo and the associated error bars as a function of the number of observed stellar rotations. In the right panel, we again plot the errors of the albedo as a function of the number of observed stellar rotations.}
\label{figure1}
\end{figure*}

        We performed this test for five different stellar rotational periods, $7$, $11$, $19$, $23$ and $26 \; \text{days}$. While the offset always stabilized around $0.999$, the rotational period $P_*$ was correct only after one entire stellar rotation, after which also the hyper-parameters reached a stable value, around $10^{-3}$ for $p_1$ and around $0.4$ for $p_2$. Since the activity pattern was the same for all the tests, the hyper-parameters did not significantly change with different rotational periods. The variation of the recovered albedo with the number of stellar rotations is shown in the left panel of Figure \ref{figure1} as functions of the number of stellar rotations. In the right panel, we show the error of the albedo as a function of the number of stellar rotations. The most evident result is that for observational lengths shorter than $P_*$ , the MCMC cannot recover the albedo. This is more evident for the $7-$ and$ \; \text{11-day}$ cases, since the associated errors reported in the right panel of Figure \ref{figure1} are very large before $1 \; P_*$ and become significantly smaller after it. For the other tests, the errors on the albedo before one entire rotation are smaller and decrease as the stellar rotational period increases. Furthermore, for the $23-$ and$ \; \text{26-day}$ cases, the left panel of Figure \ref{figure1} shows that the MCMC converges toward an albedo closer to the real one. The main explanation for this is that as the rotational period becomes longer, the rotation covers a longer time span and is described by a higher number of data. This facilitates isolating the reflected light component of the planet. 
        
        As the observations become longer, the precision in the detection of the albedo increases, stabilizing around the input value $0.3$. At between one and two stellar rotations, the albedo is compatible with the input albedo within $2 \sigma$. For more than two stellar rotations, the values are compatible to within $1 \sigma$. 
        
        As a final test, we changed the prior on the stellar rotation period and adopted a uniform distribution from $1$ to $28 \; \text{days}$. Then we applied the modified tool on a$ \; \text{19-day}$ stellar rotation simulation, with a data length of $39 \; \text{days}$. The analysis has given an albedo and a $P_*$ analogous and with a similar precision as those obtained with the restricted prior on the stellar rotation. This means that when we do not have any information about the stellar rotation, it is still possible to derive it directly from the photometric observations. For simplicity, we decided to keep the Gaussian prior on $P_*$ for all the following tests. 
        
\subsubsection{Fast rotators}

\begin{figure}
\centering
\includegraphics[width=9.1cm]{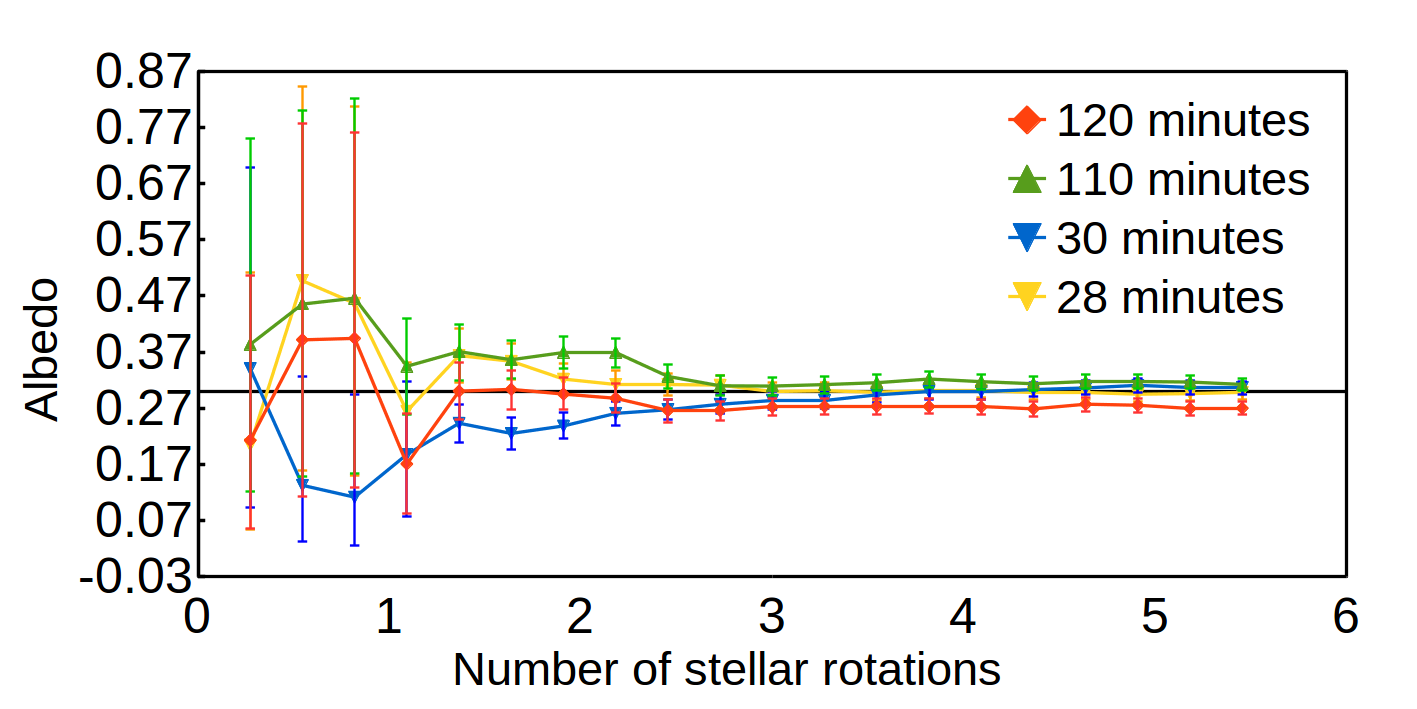}
\caption{Comparison between albedo values obtained for the$ \;\text{11-day}$ rotator and with increasing duration of the observations, but in simulations with four different timings, $120$ minutes as usual, $110$ minutes, $30$ minutes, and $28$ minutes. The $x$-axis is the number of observed stellar rotations. For all the analyzed light curves, the unmentioned input properties are the same as described in the caption of Figure \ref{figure1}.}
\label{figure2}
\end{figure}
 
        The plot of the albedo errors as a function of the number of stellar rotations in the right panel of Figure \ref{figure1} indicates a different behavior of fast and slow rotators. Fast rotators are characterized by much larger errors before one stellar rotation, and immediately after, the error decreases. In the left panel of Figure \ref{figure1}, we focus on the $7-$ and$ \; \text{11-day}$ cases. For the$ \; \text{11-day}$ rotator, the retrieved albedo value after eight orbital observation periods is almost constant and slightly overestimated with respect to the original albedo, but compatible with it within $2 \sigma$. For the$ \; \text{7-day}$ case, we can also observe a certain overestimation, with the albedo stabilizing at slightly higher than the original value of $0.3$, but the results are compatible with it within $1 \sigma$. We conclude that for fast rotators, the albedo is overestimated after long observations, but in the$ \; \text{11-day}$ tests, this trend is more significant.
        
        This overestimation might be explained with a smaller amount of data within one rotation. A higher resolution might help in determining a more reliable albedo. To demonstrate this, we generated three additional simulation sets with a rotational period of $11 \; days$, but with a cadence of $110$, $30,$ and $28$ minutes. We chose the $28$ and$\text{110-minute}$  cadence to have patterns with similar binning as the other two sets and compared them within each other. The different timing results are shown in Figure \ref{figure2}, which again reports the albedo as a function of the number of stellar rotations. The $\text{120-minute}$  timing trend is the worst case, with an underestimation of the albedo for longer observations. In the $\text{110-minute}$  case, which should be similar, the albedo is slightly overestimated, but for observations longer than $\text{two}$ stellar rotations, it is compatible within $1 \sigma$ with the inputed value. The $30$ and $\text{28-minute}$  timings are both compatible with the initial albedo after $\text{two}$ entire rotations, but with values that are much closer to the initial ones. We can conclude from this that a smaller binning helps in obtaining a better albedo and that for different binnings, there is no correlation between the number of data points and the possible underestimation or overestimation of the albedo.
         
\subsection{Variation with stellar magnitude}
\label{magnitude}       
\begin{figure}
\centering
\includegraphics[width=9.1cm]{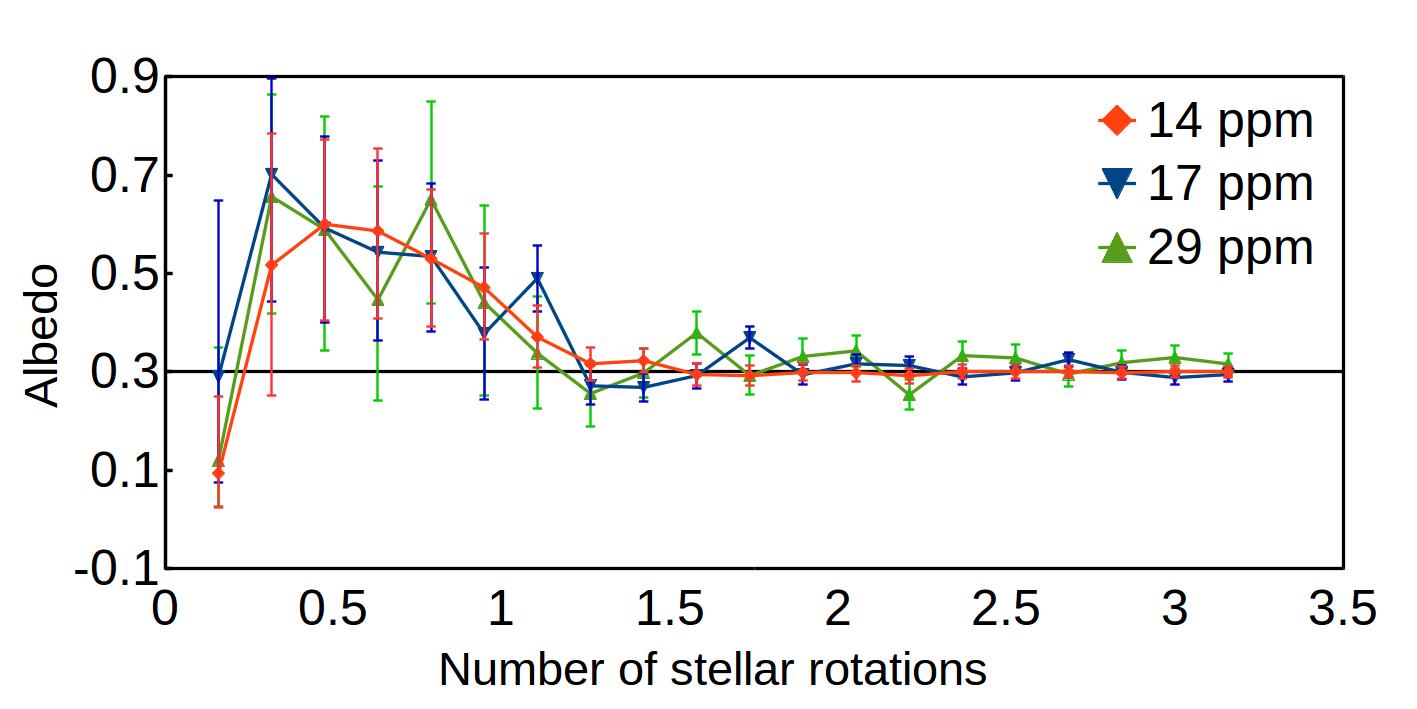}
\caption{Recovered albedo and relative error bars as a function of the number of stellar rotations for a$ \; \text{19-day}$ rotator and with three different instrumental noises, $14$, $17,$ and $29 \; ppm$ per $120$ minutes of observations. All the unmentioned input properties of the simulations are the same as reported in the caption of Figure \ref{figure1}.}
\label{figure4}
\end{figure}            
        
        As a second test, we produced two additional series of simulations for a rotation period of $19 \; days$, changing the stellar magnitude and, consequently, the instrumental noise. The adopted magnitudes and noise value are reported in Table \ref{tab1}. The resulting albedo values and error bars are reported in Figure \ref{figure4} as a function of the number of stellar rotations observed. The albedo is compatible with the input one after $20 \; days$ of observations and the error on the albedo is higher when the light curve is more noisy. All the results for $14$ and $17 \; ppm$ are compatible with the real value within $1 \sigma$. The $29 \; ppm$ case shows more oscillations, but after $2.5$ stellar rotations, it tends to stabilize on an albedo compatible with the injected one again within $1 \sigma$.
        
\begin{figure*}
\centering
\includegraphics[width=9.1cm]{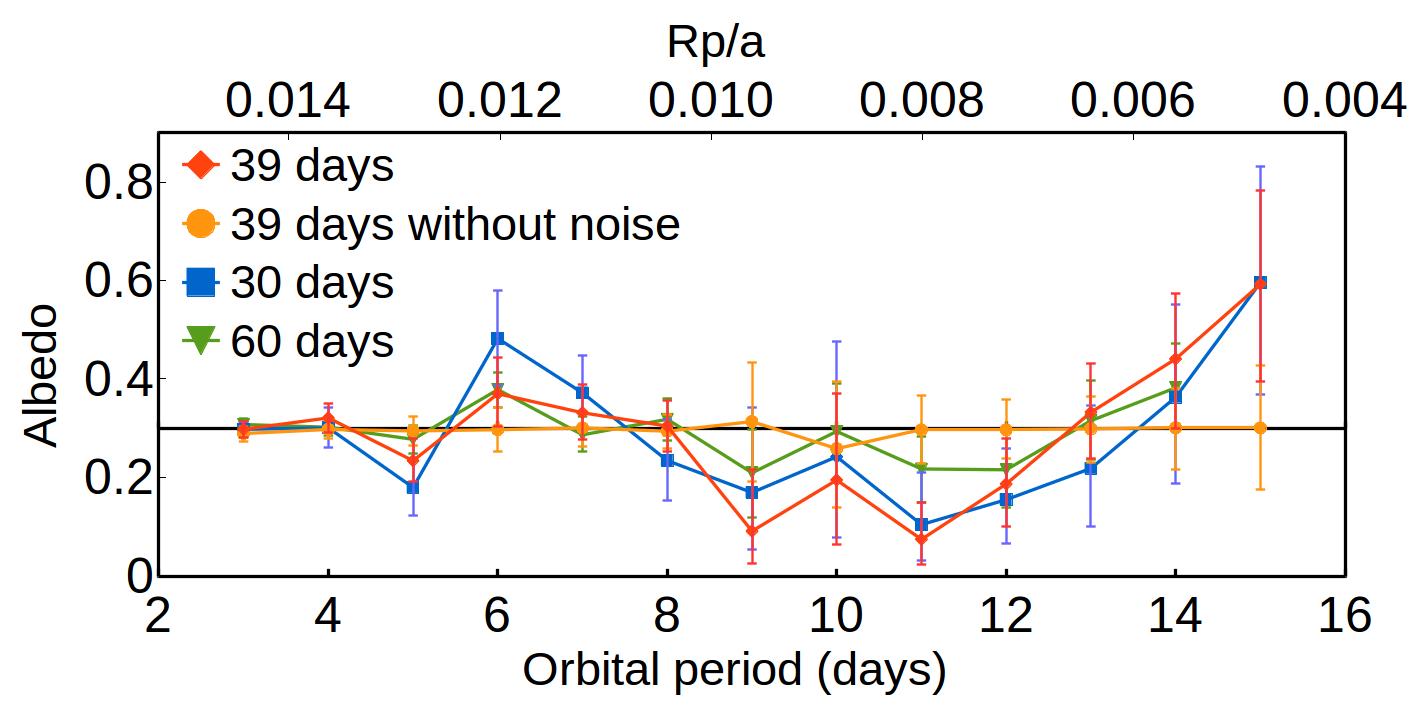}
\includegraphics[width=9.1cm]{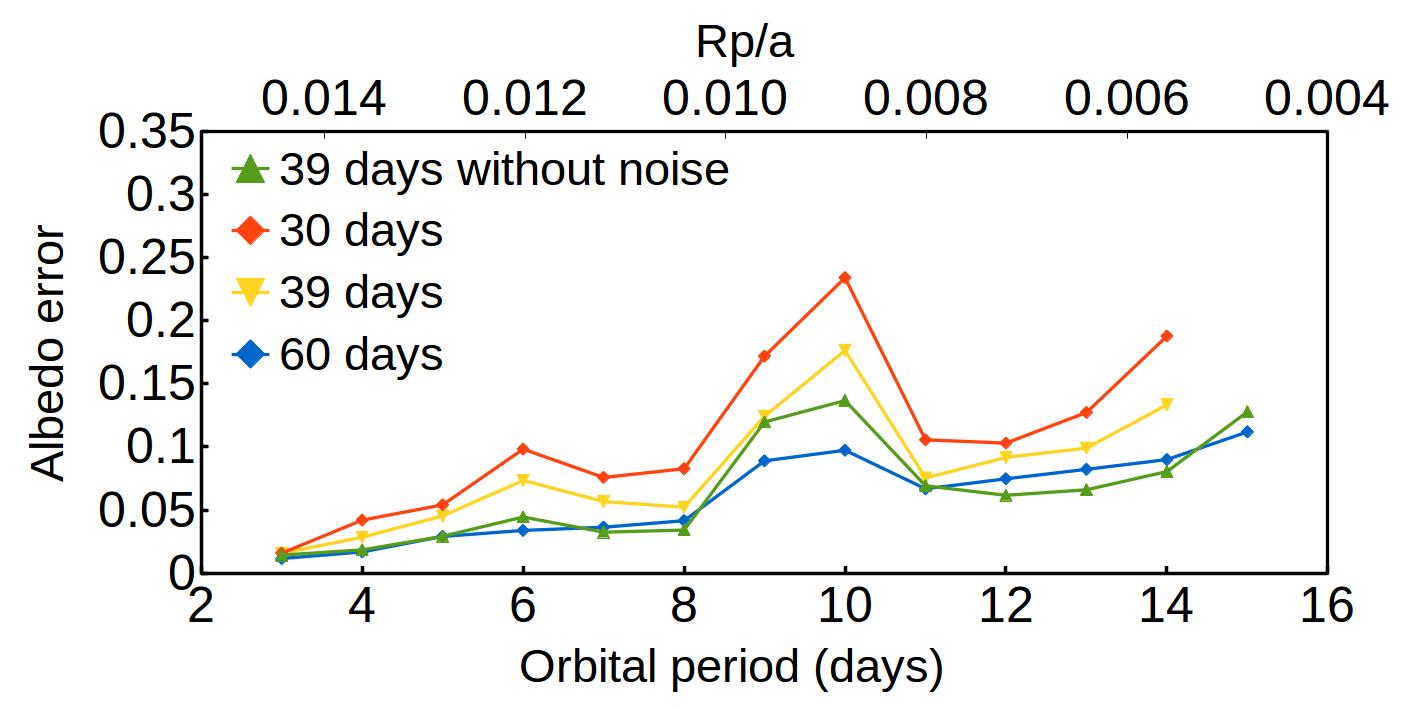}
\caption{Left: recovered albedo and relative error bars as a function of the orbital period for a$ \; \text{19-day}$ rotator for simulations with$ \; \text{39 days}$ with and without stellar activity and with $30$ and $60 \; days$ in presence of activity. Right: errors of the albedo as a function of the number of stellar rotation observed, for the simulations with $P_* = 19 \; days$ and observational lengths of $30$, $39,$ and $60 \; days$. Here we also add the error of the$ \; \text{39-day-long} $  simulation, but without stellar activity. For all the considered light curves, the input unmentioned properties are the same as in the caption of Figure \ref{figure1}. In both plots we also added the quantity $R_p/a$ as secondary horizontal axis.}
\label{figure6}
\end{figure*}

\subsection{Variation with orbital period}

\begin{figure}
\centering
\includegraphics[width=9.1cm]{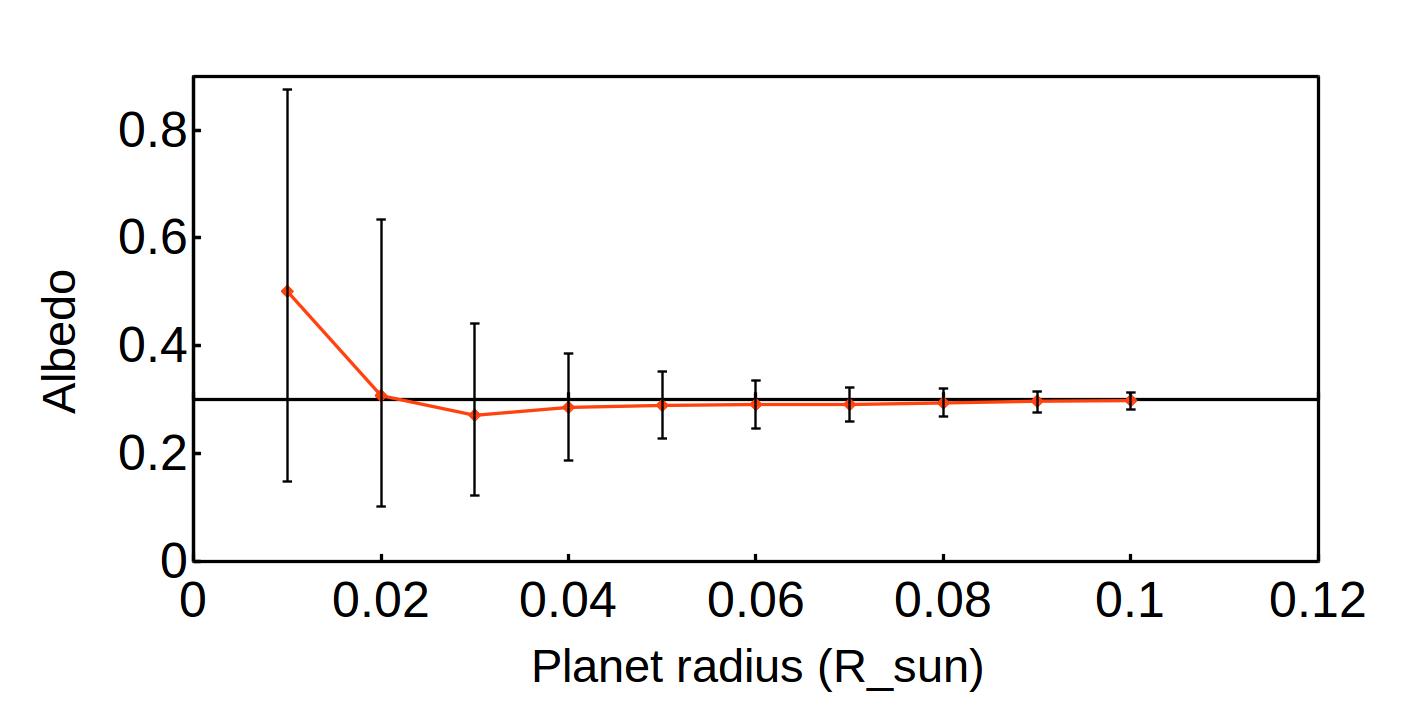}
\caption{Recovered albedo as a function of the planetary radius for$ \; \text{39-day-long}$  simulation, a stellar rotation of $19 \; days,$ and an albedo of $0.3$.}
\label{figure7}
\end{figure}    

        In this section, we test how the recovered albedo changes as the orbital period increases. An increment in orbital period determines an increment in the semi-major axis of the planet and thus a decrement of the ratio $R_p/a$, between the size of the planet and the semi-major axis of the orbit. Since the square of this ratio directly enters   Eq. \ref{amplitude} of the amplitude of the planetary phase curve, we can state that increasing the semi-major axis decreases the signal of the planet. Our objective is now to understand the limit in orbital period above which it is no longer possible to trust the albedo detection. 
        
        We fixed the stellar rotation to $19 \; days$ and varied the orbital period of the planet from $3$ to $15 \; days$ with steps of one day. We tested this for $30-$, $39-$ and$ \; \text{60-day}$ data lengths, and in Figure \ref{figure6} we report the recovered albedo as a function of the orbital period. As reference for the secondary horizontal axis, we used the quantity $R_p/a$. The worst case, with an entirely incorrect albedo for orbital periods longer than $4 \; days$, is the$ \; \text{30-day-long}$  set of data, when the $R_p/a$ is already lower than $8 \times 10^{-3}$. In the$ \; \text{39-day}$ case, the longest observations allows measuring reliable albedos up to an orbital period of $8 \; days$, corresponding to $R_p/a < 6*10^{-3}$. With $60 \; days$ of observation, the albedo is always compatible with the original one within $1 \sigma$, except for the cases with orbital periods of $9$, $11,$ and $12 \; days$, for which we can recover the albedo at $3 \sigma$.
        
        We also produced the same simulations without stellar activity and with a length of $39 \; days,$ and the results are reported together with the others in the left panel of Figure \ref{figure6}. We can observe that the albedo is always recovered, well within $1 \sigma$ until at $8 \; days$ of the orbital period, the albedo is well estimated with a small error bar. This means that for periods longer than $4 \; days$ for the$ \; \text{30-day-long}$  observations and $8 \; days$ for the $\text{39-day-long}$  observations, the white noise mainly prevents us from retrieving the albedo. For the orbital periods of $9$ and $10\; days,$ the albedo is compatible with the injected value, but as in the case of the tests in presence of white noise, the error bar is higher and the albedo is no longer exactly $0.3$. This occurs because the rotational period of the star, $19 \; days,$ is close to twice the planet period, and it determines a degeneration between the activity trend and the orbital period, which requires a higher number of observed stellar rotations to be distinguished. This effect is then amplified when we observe the trends in presence of stellar activity, where the effect of this degeneracy is also
evident for the orbital period of $11 \; days$.

        In the right panel of Figure \ref{figure6}, we report the errors associated with the albedo as a function of the orbital period, and we again use as secondary axis $R_p/a$. The$ \; \text{39-day-long}$  observation without white noise shows errors that are compatible with those for the$ \; \text{60-day}$ case until an orbital period of $8 \; days$. The $9$-, $10-,$ and$ \; \text{11-day}$ cases show larger errors in the trends with white noise, for the same reason for which the correspondent albedo increases. After this, the instrumental noise becomes predominant and the error increases. The general trend of the plot shows that as the period of the planet increases, the error of the identified albedo also increases. This occurs because the signal of the planet becomes lower and requires longer observations to be identified.

\begin{figure}
\centering
\includegraphics[width=9.1cm]{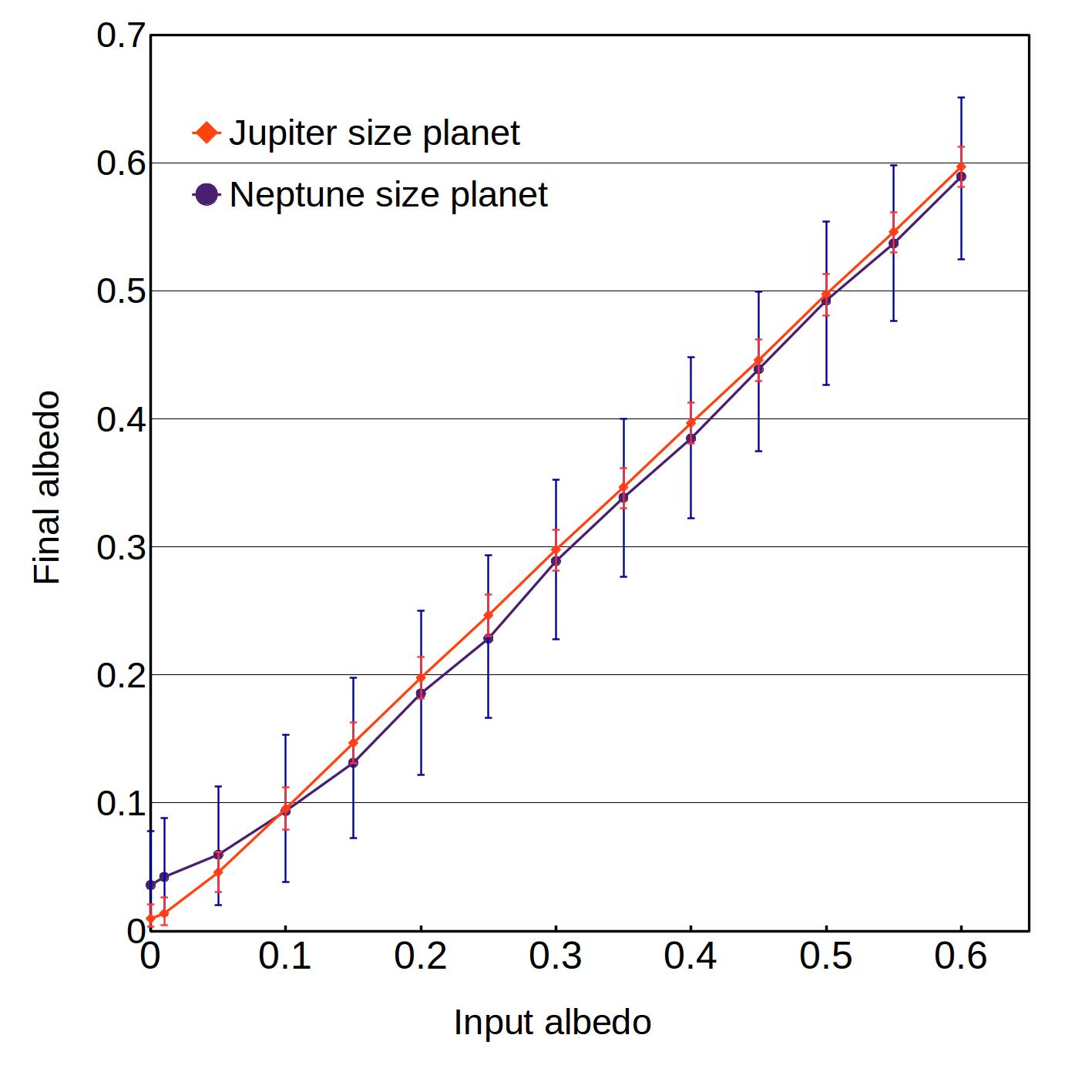}
\caption{Recovered albedo values as a function of the input values for the 39-day-long simulation and a stellar rotation of $19 \; days$. The red data points represent a Jupiter-sized planet, and the blue points show a Neptunian case. The unmentioned input properties of the simulations are the same as in the caption of Figure \ref{figure1}.}
\label{figure8}
\end{figure}                            

\subsection{Variation with planetary radius}
        As an additional test, we fixed the data length of the
observation to $39 \; days$, the albedo to $0.3,$ and the stellar rotation to $19 \; days$. Then, we generated simulations in presence of stellar activity and instrumental noise, decreasing the planetary radius from $0.1 \; R_*$ to $0.01 \; R_*$. We performed the MCMC, and the results are reported as a function of the planetary radius in Figure \ref{figure7}. We can conclude that the albedo is well estimated until a planetary radius of $0.04 \; R_*$, even if the error bars become already $0.12$ with $0.05 R_*$ planets. This shows that with high uncertainties, it is possible to detect the albedo of small Neptunes. For smaller planets, the uncertainties increase significantly and the albedo values tend to be overestimated;
they approach the center of the prior. This means that when the planets are very small, the instrumental errors are so large that the albedo posterior and prior coincide.
        
        We also decided to fix the planetary radius to $0.1 \; R_*$ and vary the albedo of $0.05$ from $0.6$ to $0$, including the $0.01$ value. The albedos obtained with the MCMC are reported as red data points as a function of the input values in Figure \ref{figure8}. In all the tests, the albedo is precisely estimated with an error bar of around $0.15$. We performed an analogous test for a Neptune-sized planet with $0.05 \; R_*$ radius, and the results are shown in the same plot as the blue data points. This time, the albedo is slightly underestimated until $A_g = 0.1$ with respect to the injected value, but it is still compatible with it within $1 \sigma$, with error bars $0.1$ long. This means
that for a fixed radius, varying the albedo does not significantly
change the level of uncertainty with which it is detected. When the albedo becomes lower than $0.1,$ the retrieved value is no
longer reliable for a Neptun-sized planet because the associated posterior distribution is a Poissonian, with a peak on $0$.
\begin{figure}
\centering
\includegraphics[width=9.1cm]{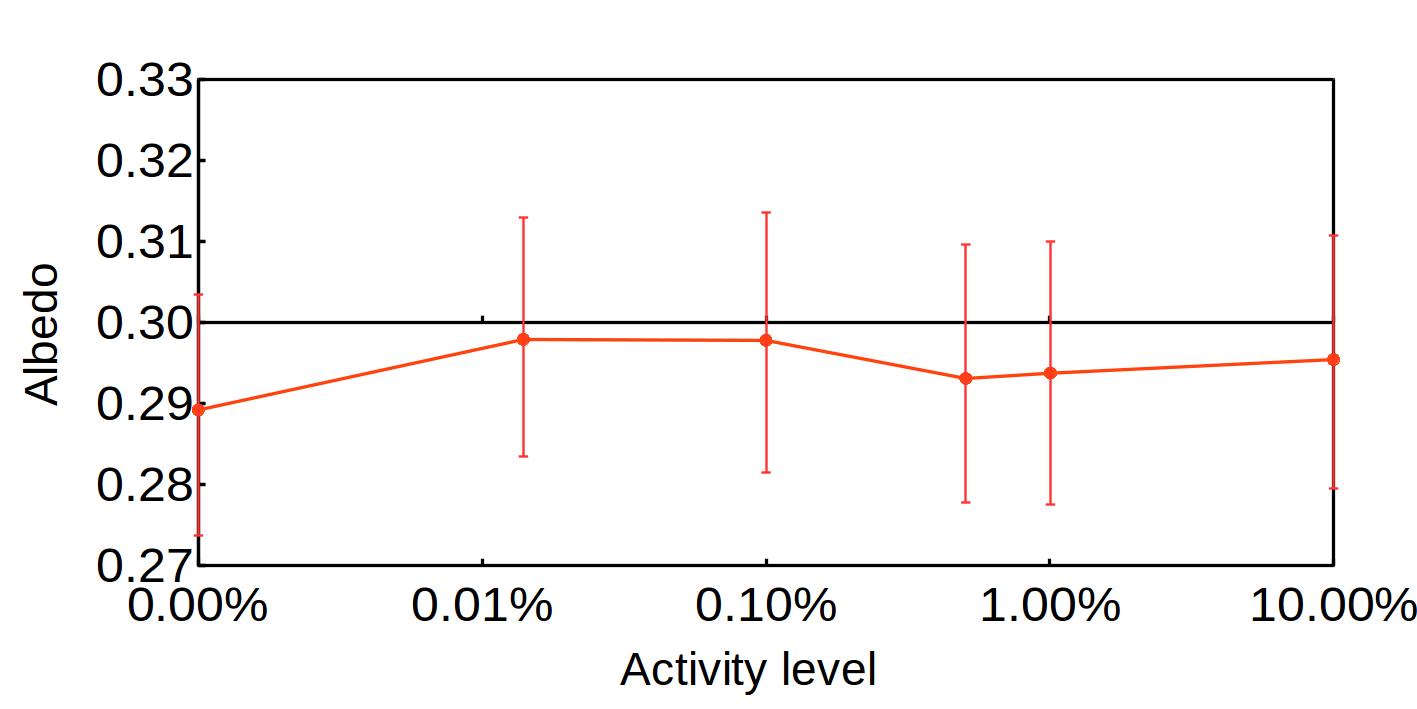}
\caption{Recovered albedo and relative error bars as a function of the activity level in percentage for$ \; \text{the
39-day-long}$ long simulation, a stellar rotation of $19 \; days$, an input albedo of $0.3,$ and a $0.1 \; R_*$ planetary radius. The unmentioned properties of the simulations are the same as in Figure \ref{figure1}. The horizontal axis is in a logarithmic scale.}
\label{figure9}
\end{figure}            
\subsection{Variation with stellar activity level}
        Finally, we tested the dependence of the recovered albedo on stellar activity level. To this aim, we fixed the observational length to $39 \; days$, the albedo to $0.3$, the planet radius again to $0.1 \; R_*$ , and the stellar rotation to $19 \; days$. Then, we varied the activity pattern by multiplying it by $100$, $10$, $5,$ and $0.1$ and adding $-99$, $-9$, $-4$, $\text{and
}0.9,$ respectively. In this way, we obtained realistic simulations with a wide range of activity levels and tested whether in all the conditions it was still possible to detect the albedo. 
        
        The results of the MCMC performed on such tests are reported in Figure \ref{figure9}, where the albedo is expressed as a function of the logarithm of the activity level. In the plot, the first point is also the test performed without stellar noise. In all the cases, the albedo was well detected, confirming also what we observed in the blind tests. The MCMC with GP can distinguish the reflected light component of a Jupiter-sized planet even in very active stars.
        
\section{Toward a complete fitting model for phase light curves}
\label{Extra}
In addition to the previous analyses, we intend to perform two additional tests that allow us to simulate more realistic observational conditions. 

In the first test, we introduced in the planetary phase curve model the previously ignored beaming effect and the ellipsoidal modulation. For these components we adopted the equations reported in \citet{Lillo-Box},
\begin{equation}
\frac{F_{ellip}}{F_*} = - \alpha_e \frac{R_*}{r} \sin ^2 i \cos 2\theta
\end{equation}
\begin{equation}
\frac{F_{beam}}{F_*} = \left(3 - \Gamma\right) \frac{K}{c} \left(\sin \theta + e \cos \omega\right)
.\end{equation}
Here, $\alpha_e$ is a factor that depends on the linear limb-darkening coefficient $u$ and on the gravity-darkening coefficient $g$,\begin{equation}
\alpha _e = 0.15 \frac{(15+u)(1+g)}{3-u}
.\end{equation}
The linear limb-darkening coefficient in this case is the parameter of the linear limb-darkening law reported in \citep{Claret1}. $K$ is the amplitude of the radial velocity,
\begin{equation}
K = 28.4 m/s \left(\frac{P}{1 \; yr}\right)^{-1/3} \frac{M_p \sin i}{M_{Jup}} \left(\frac{M_*}{M_{\odot}}\right)^{-2/3}
.\end{equation}
The $\Gamma$ factor is given by
\begin{equation}
\Gamma = \frac{e^x (3-x) - 3}{e^x - 1}
,\end{equation}
with $x = hc/k_B \lambda T_{eff}$ and $\lambda = 5750 {\AA}$ for the Kepler band.

The new light curves are given by
\begin{equation}
\label{new_model}
\frac{F_{total}}{F_*} = \frac{F_p}{F_*} + \frac{F_{beam}}{F_*} + \frac{F_{ellip}}{F_*}+ \frac{F_{*,spotted}}{F_*} + \frac{F_{noise}}{F_*}
.\end{equation}
We additionally modified the fitting tool described in section \ref{GPMCMC} by adding the two effects to the model and making the planetary mass an additional parameter. We assumed a Gaussian prior for the mass, with an average $1$ Jupiter mass and standard deviation $0.02$. Then, we modeled a stellar light curve as in equation \ref{new_model}, considering the case of a Jupiter-sized planet, with an albedo of $0.3$ and with $1$ Jupiter mass, orbiting a Sun-like star with a rotational period of $19 \; days$. All the other properties of the star are the same as in Table \ref{tab3}, while for the planet, we used the
parameters listed in Table \ref{tab6}. In addition, for the linear limb-darkening coefficient, $u$, and the gravity-darkening coefficient, $g$, we used typical values for Sun-like stars from \citet{Claret1} ($u =  0.6230$, $g = 0.3456$). 
        
        With this setup, we ran the MCMC with the same parameters as before. The recovered albedo is $0.296$, close to the one determined without a beaming effect and ellipsoidal modulation and with exactly the same uncertainty, $0.016$. This suggests that adding the two additional components of the planetary phase curve does not change the precision with which we retrieve the reflected light component of the planet. The posterior distribution of the mass is the same as the input prior, with the uncertainty around the median equal to the input value of $0.015$. This means that the data are not able to constrain the planet mass. The amplitudes of the beaming effect and the ellipsoidal modulation are, indeed, close to $2 \; ppm$ for a $1$ Jupiter-mass planet, which is much lower than the level of instrumental noise we introduced in the simulation. Therefore, in this analysis we cannot distinguish the two signals from the noise present in the light curve.

In the second test, we introduced an uncertainty around the planet radius and the semi-major axis comparable to the typical uncertainties with which they are retrieved from the transit fit. This correctly accounts for the uncertainties of the parameters that can be estimated from a transit. We thus increased the number of parameters of the MCMC, adding the planet radius and the semi-major axis, both in units of stellar radii. We used Gaussian priors with a standard deviation pf $0.005$ for the planet radius and $0.05$ for the semi-major axis \citep[the typical uncertainties for ground-based transits of hot Jupiters around bright stars, see][]{Turner}. The model for the phase curve remained the same, the sum of the reflected light with the beaming effect and the ellipsoidal modulation, and we fit the same light curve as above. The final value of the albedo is $0.302$, slightly different from the valuee obtained when we had fixed these parameters, which was $0.296$, but the uncertainty is $0.040$, higher than the previous value of $0.015$. This is a consequence of the fact that inside the reflected light component, the radius and the semi-major axis both appear, together with the albedo, in the form of $(R_p/a)^2$. Applying the theory of error propagation, we calculate that the ratio $R_p/a$ has itself a standard deviation of $0.0006$, which decreases the precision with which we can retrieve the albedo. Moreover, the posterior distributions on the planet radius and the semi-major axis are close to their priors, therefore we cannot improve our knowledge on them with the reflected light component alone, and even though the semi-major axis appears in the beaming effect and the ellipsoidal modulation, we already know that these are too small to be isolated from the instrumental noise without properly knowing the mass. The error of the recovered albedo will always depend on the uncertainties of the two physical quantities $R_p$ and $a$.
        
\section{Test with CHEOPS gaps}
\label{gaps}
\label{holes}
        
\begin{figure*}
\centering
\includegraphics[width=20cm]{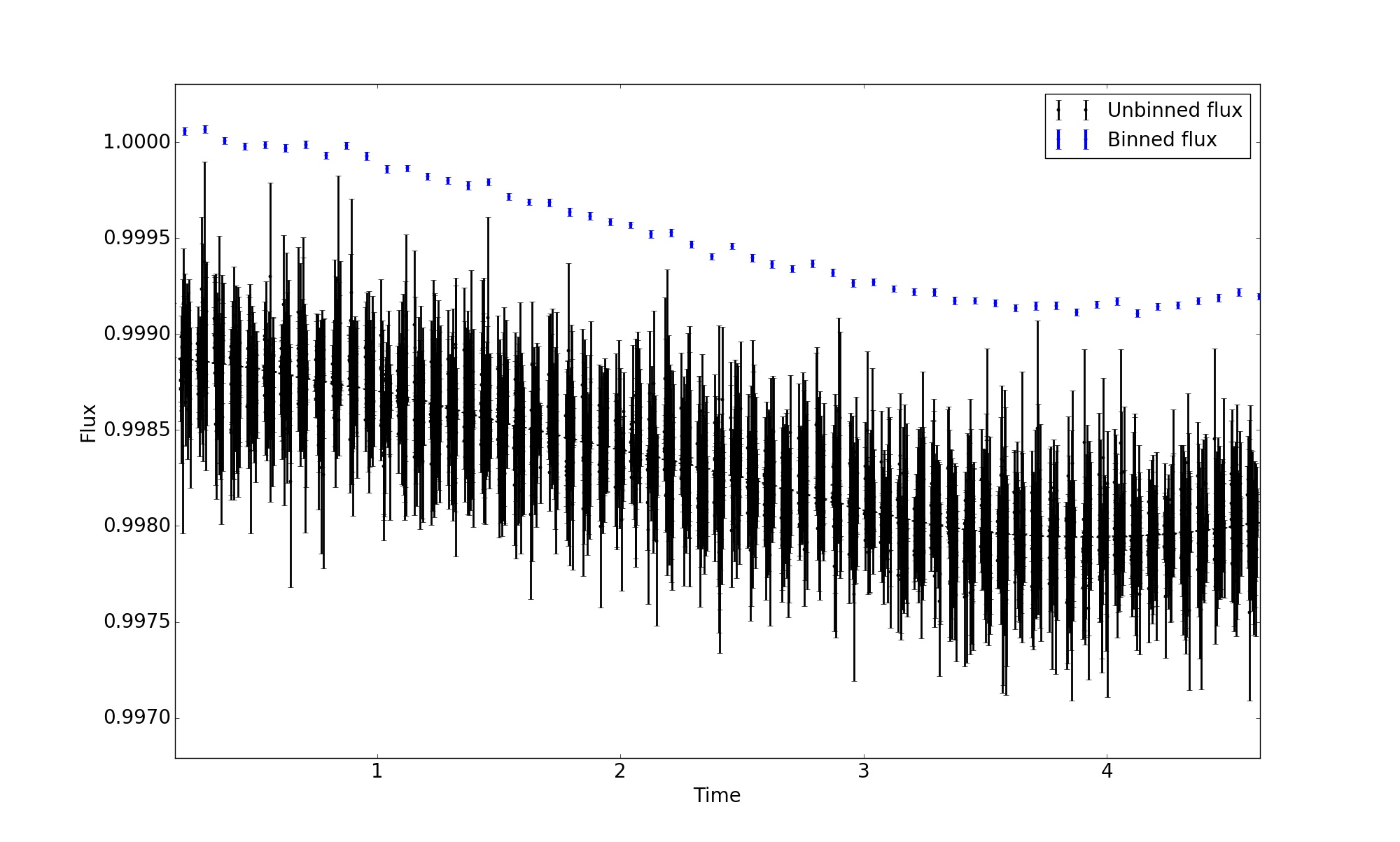}
\caption{Simulation of stellar light curve in presence of gaps and with a timing of $1$ minute. In black we report the generated simulation, and in red the binned simulation. The gaps only cover some minutes. Time is expressed in days.}
\label{figure10}
\end{figure*}           
        
        In more realistic conditions, as anticipated in Section \ref{simulations}, the satellite CHEOPS will not be able to perform continuous observations. During an observation set, there will be small gaps in the data. To understand whether this could represent a problem in the detection of the albedo, we simulated a light curve with a $40\%$ gap as expected for an average CHEOPS observation. These gaps are distributed in time in a realistic way (courtesy of M. Lendl). This data-set has a cadence of one minute and instrumental noise of $155 \; ppm$ per exposure, as predicted for a 6.5 magnitude star. We adopted the stellar activity pattern in Table \ref{tab6}. The planet albedo is $0.3$ and its radius is $0.1 \;R_*$. Its other properties are listed in Table \ref{tab7}. 

        After this, we binned the simulation to have a data point every two hours. In Figure \ref{figure10} we report the unbinned simulation, with the flux as a function of time. The black line shows the overall light curve, and the red line plots the binned curve. With this plot, we can observe that the gaps just cover a time span of some minutes (a small fraction of the orbital period of $100$ minutes). Consequently, the net effect of the binning procedure is that the error of each binned data point is larger by $40\%$ than those for$ \; \text{two-hour}$ timing data without gaps. We applied our analysis tool on a$ \; \text{39-day-long}$  binned simulation. The resulting albedo was $0.2898^{[+0.015]}_{[-0.015]}$, compatible within $1 \sigma$ with the real value of $0.3$, but still lower than the one obtained with the same initial parameters and without gaps. The stellar rotation is $P_* = 19.001^{[+0.003]}_{[-0.003]} \; days$, compatible with the input one of $19 \; days$ within $2 \sigma$. The increment in noise only slightly affected the albedo. Thus, the gaps in CHEOPS observations will not significantly change the reliability of the data analysis.    
        
\begin{figure*}
\label{Kepler_star}
\centering
\includegraphics[width=20cm]{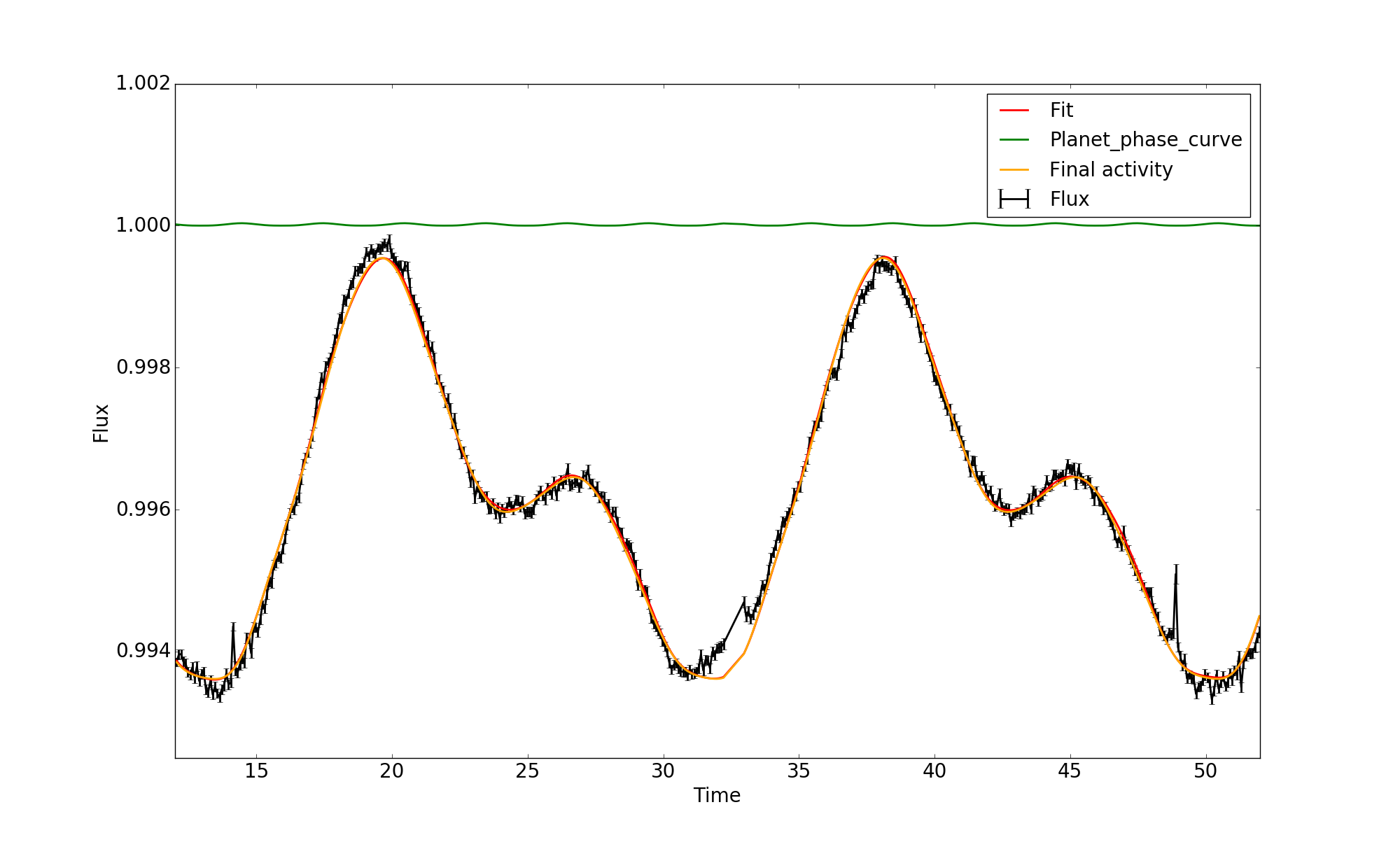}
\caption{Extraction of the 12th quarter of Kepler observations for the star KIC 3643000 after adding a planet and a two-hour binning. The black error bars represent the data, the red line shows the fit, the orange line show the identified stellar activity, and the green line plots the planetary phase curve shifted by 1. The planet phase modulation is built with an albedo of $0.3$, a planetary radius $R_p = 0.1 \; R*$ , and the same properties as in Table \ref{tab7}.}
\end{figure*}
        
\section{Tests on real data: Kepler-7 and  KIC 3643000} 
\label{Kepler_star_sec}
        To verify that our code could work in more realistic conditions, we applied it to Kepler-7, a star with a rotation period of $16.7 \; days$ that is orbited by a $1.6 R_{*}$ planet. We extracted the entire 10th quarter of Kepler observations and binned the data to two hours. A visual inspection of the light curve easily showed spot evolution, which suggested that the periodic kernel is not the proper model for describing the stellar activity. After we applied our fitting tool, we obtained an albedo
of $0.36$, very close to the value $0.35$ reported in the literature \citep{Angerhausen}, while the stellar activity signal is not reliably recovered. The rotation period of the star is $15.7 \; days$ instead of $16.7$ days. The other hyper-parameters are also physically poorly constrained and the retrieved activity is not periodical, as it should be due to the adopted kernel. This occurs because the Kepler-7 activity pattern is not strictly periodic and shows evidence of spot evolution.
        
        Since the GP in our code does not model the spot evolution, we selected a Kepler star that visually showed a periodic variability, in agreement with the initial conditions imposed by our tool. To this star we added a planetary phase curve modeled as described in section \ref{simulation_planet}. The planet had the properties reported in table \ref{tab7}, as well as $R_p = 0.1 \; R_*$, an orbital period of $3 \; days$ and an albedo
of $0.3$. The selected star has the Kepler identifier KIC 3643000, a rotationally variable star with a rotational period $P_* = 18.94 \; days$ that is extracted from the McQuillan catalog of stars \citep{McQuillan}. In particular, we used a portion of the light curve of the star KIC 3643000 that was retrieved during
the 12th quarter of Kepler observations, in which the activity features do not change significantly. In this way, the periodic kernel of the GP could be more realistic for analyzing the selected data. After adding the planet, we binned the data to have the same sampling as before. The final level of noise we obtained was about $39 \; ppm$, higher than the value we used in the tests
so far.
        
        Then, we applied the fitting procedure to the data and retrieved $A_g = 0.26 \pm 0.11$ and $P_* = 18.4818 ^{+0.0071} _{-0.00069}$. With respect to the simulated data, the albedo error bar is twice as large. This might be due to the instrumental error, but there might be another explanation. A possibly not completely constrained stellar rotational period (which is lower than the period reported in the literature) might have changed the information on the albedo. Following the results of section \ref{magnitude} and with the opportunity of observing more than the two stellar rotations selected in this case, the precision might have increased significantly. We concluded that we were able to retrieve the albedo within $1 \sigma$ and with a reasonable error bar. Therefore, we consider our method reliable. In figure \ref{Kepler_star} we report the modified light curve of the star in black, which includes the planetary modulation, in red we show the fit obtained at the end of the performed MCMC, in orange the activity identified, and in green the planetary phase curve modulation, shifted up by 1.
\label{simulation}

\section{Conclusions}
We have analyzed the effect of stellar activity and instrumental noise on the detection of the planetary albedo by measuring the planet phase curve from a photometric light curve. To this end, we built simulations of the stellar light curve that included the reflected light modulation of a planet as a function of the geometric albedo, a stellar activity pattern, modeled with SOAP-T  \citep{Oshagh1}, and instrumental noise. As noise we adopted the predicted noise for the future spacecraft CHEOPS, which will perform photometric observations of bright stars.

        To analyze the detectability of the planetary albedo, we used an MCMC with a GP to model the stellar activity pattern. Our MCMC adopts as parameters the albedo, an offset that represents the average of the stellar activity pattern, and the hyper-parameters of the GP, which include the stellar rotation period. We showed using blind tests that our model can correctly recover the albedo for a wide range of stellar activity patterns.
        
        We tested our method on real Kepler data of an active star. We also explored the detectability of the albedo both in presence of the stellar activity and the instrumental noise. We conclude the following:
        \begin{itemize}
        \item For observations shorter than one full stellar rotation, our method cannot recover the rotation period of the star. This prevents a correct estimation of the albedo.
        \item The procedure works better for slow rotators. For fast rotators, a smaller sampling is required to obtain more reliable results.
        \item For the brightest stars, CHEOPS can measure the albedo of the exoplanets with an orbital period  maximum of $13 \; days$ .
        \item For a noise level of $155 \; ppm$ per minute (as achieved for the brightest objects with CHEOPS, $magV \sim 6.5$), we have the opportunity of detecting the albedo of the smallest Neptunes.
        \item Adding the beaming effect and the ellipsoidal modulation does not change the precision with which we estimate the albedo.
        \item Introducing uncertainties on the planet radius and the semi-major axis increases the uncertainties on the detected albedo.
        \item Our method works also in presence of gaps in the observations, as are expected for CHEOPS.
        \item Our method works also when we adopt as stellar activity that of a Kepler star, even if the higher instrumental noise increases the associated error bar.
        \end{itemize}
        
        In summary, with CHEOPS noise, for shortperiod planets and an observation covering more than $\text{one}$ stellar rotation, we can distinguish the albedo from the stellar activity even when the activity level is very high. This result is valid for any photometric follow-up performed on stars with orbiting planets because the precision in the albedo detection mainly depends on the level of instrumental noise when the stellar activity has been well modeled with the GP. Since we have not introduced the transit feature in our analysis tool, our phase curve analysis suggests that it is possible to detect the phase light curves of non-transiting exoplanets in presence of stellar activity and instrumental noise \citep{Crossfield}. It is our intention to improve our tool by adding the beaming effect and the ellipsoidal modulation to the model, by taking into account the spot evolution, and by introducing the other planetary properties as parameters in the MCMC.

\begin{acknowledgements}
This work was supported by Fundação para a Ciência e Tecnologia (FCT) through national fundsand by FEDER through COMPETE2020 by these grants UID/FIS/04434/2013 \& POCI-01-0145-FEDER-007672, PTDC/FIS-AST/1526/2014 \& POCI-01-0145-FEDER-016886 and PTDC/FIS-AST/7073/2014 \& POCI-01-0145-FEDER-016880. N.C.S., S.G.S.and S.C.C.B. also acknowledge support from FCT through Investigador FCT contracts nºs IF/00169/2012/CP0150/CT0002, IF/00028/2014/CP1215/CT0002 and IF/01312/2014/CP1215/CT0004. L.M.S. and J.P.F. also acknowledge support by the fellowship SFRH/BD/120518/2016 and SFRH/BD/93848/2013 funded by FCT (Portugal) and POPH/FSE (EC). M.O. acknowledges research funding from the Deutsche Forschungsgemeinschft (DFG, German Research Foundation)-OS 508/1-1. M.O also acknowledges the support of COST Action TD1308 through STSM grant with reference Number:STSM-D1308-050217-081659. We would like to thank the anonymous referee for insightful suggestions that significantly clarified this paper.
\end{acknowledgements}

\bibliography{bibliography_albedo}
\end{document}